\documentclass[twocolumn,english,aps, superscriptaddress]{revtex4}
\usepackage[T1]{fontenc}
\usepackage[latin9]{inputenc}
\usepackage{amsmath}
\usepackage{graphicx}
\usepackage{amssymb}
\usepackage{babel}

\begin{document}

\title{Criteria for generalized macroscopic and mesoscopic quantum coherence}

\author{E. G. Cavalcanti}

\affiliation{Centre for Quantum Dynamics, Griffith University, Brisbane, Australia}

\affiliation{ARC Centre of Excellence for Quantum-Atom Optics, The University
of Queensland, Brisbane, Australia}

\author{M. D. Reid}

\affiliation{ARC Centre of Excellence for Quantum-Atom Optics, The University
of Queensland, Brisbane, Australia}

\date{\today{}}

\begin{abstract}
We consider macroscopic, mesoscopic and `$S$-scopic' quantum superpositions
of eigenstates of an observable, and develop some signatures for their
existence. We define the extent, or size $S$ of a superposition,
with respect to an observable $\hat{x}$, as being the range of outcomes
of $\hat{x}$ predicted by that superposition. Such superpositions
are referred to as generalized $S$-scopic superpositions to distinguish
them from the extreme superpositions that superpose only the two states
that have a difference $S$ in their prediction for the observable.
We also consider generalized $S$-scopic superpositions of coherent
states. We explore the constraints that are placed on the statistics
if we suppose a system to be described by mixtures of superpositions
that are restricted in size. In this way we arrive at experimental
criteria that are sufficient to deduce the existence of a generalized
$S$-scopic superposition. The signatures developed are useful where
one is able to demonstrate a degree of squeezing. We also discuss
how the signatures enable a new type of Einstein-Podolsky-Rosen gedanken
experiment.
\end{abstract}
\maketitle

\section{introduction}

Since Schrödinger's seminal essay of 1935 \citep{schroedinger 1935},
in which he introduced his famous cat paradox, there has been a great
deal of interest and debate on the subject of the existence of a superposition
of two macroscopically distinguishable states. This issue is closely
related to the so-called \emph{measurement problem} \citep{zurek}.
Some attempts to solve this problem, such as that of Ghirardi, Rimini,
Weber and Pearle \citep{measuremnttheories}, introduce modified dynamics
that cause a collapse of the wave function, effectively limiting the
size of allowed superpositions.

It thus becomes relevant to determine whether a superposition of states
with a certain level of distinguishability can exist experimentally
\citep{testsmt}. Evidence \citep{schexp,philcats} for quantum superpositions
of two distinguishable states has been put forward for a range of
different physical systems including SQUIDs, trapped ions, optical
photons and photons in microwave high-Q cavities. Signatures for the
size of superpositions have been discussed by Leggett \citep{disleg}
and, more recently, by Korsbakken et al \citep{brig}. Theoretical
work suggests that the generation of a superposition of two truly
macroscopically distinct states will be greatly hindered by decoherence
\citep{decoh,zureck}. 

Recently \citep{macsupPRL}, we suggested to broaden the concept of
detection of macroscopic superpositions, by focusing on signatures
that confirm, for some experimental instance, a failure of microscopic/mesoscopic
superpositions to predict the measured statistics. This approach is
applicable to a broader range of experimental situations based on
macroscopic systems, where there would be a macroscopic range of outcomes
for some observable, but not necessarily just two that are macroscopically
distinct. Recent work by Marquardt et al \citep{marqu} reports experimental
application of this approach. 

The paradigmatic example \citep{legg,leggcont,schexp,philcats} of
a macroscopic superposition involves two states $\psi_{+}$ and $\psi_{-}$,
macroscopically distinct in the sense that the respective outcomes
of a measurement $\hat{x}$ fall into regions of outcome domain, denoted
$+$ and $-$, that are macroscopically different. We argue in \citep{macsupPRL}
that a superposition of type \begin{equation}
\psi_{+}+\psi_{0}+\psi_{-},\label{eq:mmsupgen}\end{equation}
that involves a range of states but with only some pairs (in this
case $\psi_{+}$ and $\psi_{-}$) macroscopically distinct must also
be considered a type of macroscopic superposition (we call these \emph{generalized
macroscopic superpositions}), in the sense that it displays a nonzero
off-diagonal density matrix element $\langle\psi_{+}|\rho|\psi_{-}\rangle$
connecting two macroscopically distinct states, and hence cannot be
constructed from microscopic superpositions of the basis states of
$\hat{x}$. Such superpositions \citep{bellspin,drummhighspin,peres,reiddemartbell}
are predicted to be generated in certain key macroscopic experiments,
that have confirmed continuous-variable \citep{eprexp,Zhang,silber,bowen,elizclau,schori,bowen pol,cornpol,elizcoldatoms,elizcoldsq,squresultexp}
squeezing and entanglement, spin squeezing and entanglement of atomic
ensembles \citep{atomsq}, and entanglement and violations of Bell
inequalities for discrete measurements on multi-photon systems \citep{demart1,multi-ent1,spinexp}.

In this paper, we expand on our previous work \citep{macsupPRL} and
derive new criteria for the detection of the generalized macroscopic
(or $S$-scopic) superpositions using continuous variable measurements.
These criteria confirm that a macroscopic system cannot be described
as any mixture of only microscopic (or $s$-scopic, where $s<S$)
quantum superpositions of eigenstates of $\hat{x}$. We show how to
apply the criteria to detect generalized $S$-scopic superpositions
in squeezed and entangled states that are of experimental interest. 

The generalized macroscopic superpositions still hold interest from
the point of view of Schrödinger's discussion \citep{schroedinger 1935}
of the apparent incompatibility of quantum mechanics with macroscopic
realism. This is so because such superpositions cannot be represented
as a mixture of states which give outcomes for $\hat{x}$ that always
correspond to one or other (or neither) of the macroscopically distinct
regions $+$ and $-$. The quantum mechanical paradoxes associated
with the generalized macroscopic superposition (\ref{eq:mmsupgen})
have been discussed in previous papers \citep{macsupPRL,bellspin,drummhighspin,macropara,newjournmidopt}.

The criteria derived in this paper take the form of inequalities.
Their derivation utilizes the uncertainty principle and the assumption
of certain types of mixtures. In this respect they are similar to
criteria for inseparability that have been derived by Duan et al \citep{inseplur}
and Hofmann and Takeuchi \citep{hof}. Rather than testing for failure
of separable states, however, they test for failure of a phase space
{}``macroscopic separability'', where it is assumed that a system
is always in a mixture (never a superposition) of macroscopically
separated states.

We will in this paper note that one can be more general in the derivation
of the inequalities, adopting the approach of Leggett and Garg \citep{legg}
to define a macroscopic reality without reference to any quantum concepts.
One may consider a whole class of theories, which we refer to as the
\emph{minimum uncertainty theories} (MUTs) and to which quantum mechanics
belongs, for which the uncertainty relations hold and the inequalities
therefore follow, based on this macroscopic reality. The experimental
confirmation of violation of these inequalities will then lead to
demonstration of a new type of Einstein-Podolsky-Rosen argument (or
{}``paradox'') \citep{epr}, in which the inconsistency of a type
of macroscopic ($S$-scopic) reality with the completeness of quantum
mechanics is revealed \citep{macsupPRL,macropara}. A direct analogy
exists with the original EPR argument, which is a demonstration of
the incompatibility of local realism with the completeness of quantum
mechanics \citep{mdrepr,wiseepr,eprreview}.

\section{Generalized $S$-scopic Coherence \label{sec:Generalized-S-scopic-Coherence}}

We introduce in this Section the concept of a generalized $S$-scopic
coherence \citep{macsupPRL}, which we define in terms of failure
of certain types of mixtures. In the next Section, we link this concept
to that of the generalized $S$-scopic superpositions (\ref{eq:mmsupgen}).

We consider a system which is in a statistical mixture of two component
states. For example, if one attributes probabilities $\wp_{1}$ and
$\wp_{2}$ to underlying quantum states $\rho_{1}$ and $\rho_{2}$,
respectively (where $\rho_{i}$ denotes a quantum density operator),
then the state of the system will be described as a mixture, which
in quantum mechanics is represented as \begin{equation}
\rho=\wp_{1}\rho_{1}+\wp_{2}\rho_{2}.\label{eq:QM mixture}\end{equation}
This can be interpreted as \char`\"{}the state is \emph{either} $\rho_{1}$
with probability $\wp_{1}$, \emph{or} $\rho_{2}$ with probability
$\wp_{2}$.\char`\"{} The probability for an outcome $x$ of any measurable
physical quantity $\hat{x}$ can be written, for a mixture of the
type (\ref{eq:QM mixture}), as\begin{equation}
P(x)=\wp_{1}P_{1}(x)+\wp_{2}P_{2}(x),\label{eq:probmacro}\end{equation}
where $P_{i}(x)$ ($i=1,2$) is the probability distribution of $x$
in the state $\rho_{i}$. 

More generally, in any physical theory, the specification of a state
$\rho$ (where here $\rho$ is just a symbol to denote the state,
but not necessarily a density matrix) fully specifies the probabilities
of outcomes of all experiments that can be performed on the system.
If we then have with probability $\wp_{1}$ a state $\rho_{1}$ which
predicts for each observable $\hat{x}$ a probability distribution
$P_{1}(x)$ and with probability $\wp_{2}$ a second state which predicts
$P_{2}(x)$, then the probability distribution for any observable
$\hat{x}$ given such mixture is of the form \eqref{eq:probmacro}.
The concept of coherence can now be introduced. 

\begin{quote}
\textbf{Definition 1:} \emph{The state of a physical system displays}
\textbf{coherence}\emph{ between two outcomes $x_{1}$ and $x_{2}$
of an observable $\hat{x}$ if and only if the state $\rho$ of the
system cannot be considered a statistical mixture of some underlying
states $\rho_{1}$ and $\rho_{2}$, where $\rho_{1}$ assigns probability
zero for $x_{2}$ and $\rho_{2}$ assigns probability zero for $x_{1}$.} 
\end{quote}
This definition is independent of quantum mechanics. Within quantum
mechanics it implies that the quantum density matrix representing
the system cannot be decomposed in the form (\ref{eq:QM mixture}).
Thus, for example, $\rho=\wp_{+}|\psi_{+}\rangle\langle\psi_{+}|+\wp_{-}|\psi_{-}\rangle\langle\psi_{-}|$
where $|\psi_{\pm}\rangle=[|x_{1}\rangle\pm|x_{2}\rangle]/\sqrt{2}$
does not display coherence between $x_{1}$ and $x_{2}$ because it
can be rewritten to satisfy (\ref{eq:QM mixture}). The definition
will allow a state to be said to have coherence between $x_{1}$ and
$x_{2}$ if and only if there is no possible ensemble decomposition
of that state which allows an interpretation as a mixture (\ref{eq:QM mixture}),
so that the system cannot be regarded as being in one or other of
the states that can generate at most one of $x_{1}$ or $x_{2}$. 

We next define the concept of \emph{generalized $S$-scopic coherence}. 

\begin{quote}
\textbf{Definition 2:} \emph{We say that the state displays }\textbf{generalised
$S$-scopic coherence}\emph{ if and only if there exist $x_{1}$ and
$x_{2}$ with $x_{2}-x_{1}\geq S$ (we take $x_{2}>x_{1}$), such
that $\rho$ displays coherence between some outcomes $x\leq x_{1}$
and $x\geq x_{2}$. This coherence will be said to be }\textbf{macroscopic}\emph{
when $S$ is macroscopic.} 
\end{quote}
If there is \emph{no} generalized\emph{ }$S$-scopic coherence, then
the system can be described as a mixture (\ref{eq:QM mixture}) where
now states $\rho_{1}$ and $\rho_{2}$ assign nonzero probability
only for $x<x_{2}$ and $x>x_{1}$ respectively. This situation is
depicted in Fig. 1.

\begin{figure}
\includegraphics[scale=0.3]{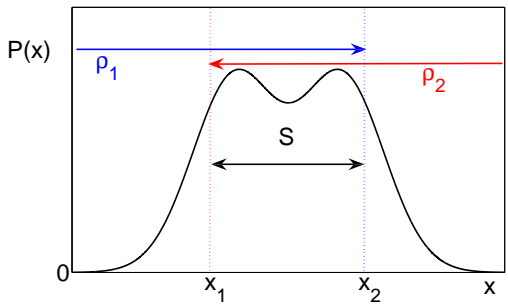}

\caption{Probability distribution for outcomes $x$ of measurement $\hat{x}$.
If $x_{1}$ and $x_{2}$ are macroscopically separated, then we might
expect the system to be described as the mixture (\ref{eq:QM mixture}),
where $\rho_{1}$ encompasses outcomes $x<x_{2}$, and $\rho_{2}$
encompasses outcomes $x>x_{1}$. This means an absence of generalized
macroscopic coherence, as defined in Sec. II.}

\end{figure}

An important clarification is needed at this point. It is clearly
a vague matter to determine when $S$ is macroscopic. What is important
is that we are able to push the boundaries of experimental demonstrations
of $S$-scopic coherence to larger values of $S$. We will keep the
simpler terminology, but the reader might want to understand \emph{macroscopic}
as \emph{S-scopic} throughout the text\emph{.}

Generalized macroscopic coherence amounts to a loss of what we will
call a \emph{generalized macroscopic reality}. The simpler form of
macroscopic reality that involves only two states macroscopically
distinct has been discussed extensively by Leggett \citep{legg,leggcont}.
This simpler case would be applicable to the situation of Fig. 1 if
there were zero probability for result in the intermediate region
$x_{1}<x<x_{2}$. Macroscopic reality in this simpler situation means
that the system must be in one or other of two macroscopically distinct
states, $\rho_{1}$ and $\rho_{2}$, that predict outcomes in regions
$x\leq x_{1}$ and $x\geq x_{2}$, respectively. The term {}``macroscopic
reality'' is used \citep{legg} because the definition precludes
that the system can be in a superposition of two macroscopically distinct
states, prior to measurement. \emph{Generalized macroscopic reality}
applies to the broader situation, where probabilities for outcomes
$x_{1}<x<x_{2}$ are not zero, and means that where we have two macroscopically
separated outcomes $x_{1}$ and $x_{2}$, the system can be interpreted
as being in one or other of two states $\rho_{1}$ and $\rho_{2}$,
that can predict \emph{at most} one of $x_{1}$ or $x_{2}$. Again,
the term macroscopic reality is used, because this definition precludes
that the system is a superposition of two states that can give macroscopically
separated outcomes $x_{1}$ and $x_{2},$ respectively. 

We note that Leggett and Garg \citep{legg} define a macroscopic reality
in which they do not restrict to quantum states $\rho_{1}$ and $\rho_{2}$,
but allow for a more general class of theories where $\rho_{1}$ and
$\rho_{2}$ can be hidden variable states of the type considered by
Bell \citep{bell}. Such states are not restricted by the uncertainty
relation that would apply to each quantum state, and hence the assumption
of macroscopic reality as applied to these theories would not lead
to the inequalities we derive in this paper. This point will be discussed
in Sec. IV, but the reader should note that the definition of $S$-scopic
coherence within quantum mechanics means that $\rho_{1}$ and $\rho_{2}$
are quantum states.

\section{Generalized macroscopic and $S$-scopic quantum Superpositions }

We now link the definition of generalized macroscopic coherence to
the definition of generalized macroscopic superposition states \citep{macsupPRL}.
Generally we can express $\rho$ as a mixture of pure states $|\psi_{i}\rangle$.
Thus\begin{equation}
\rho=\sum_{i}\wp_{i}|\psi_{i}\rangle\langle\psi_{i}|,\label{eq:mixture}\end{equation}
where we can expand each $|\psi_{i}\rangle$ in terms of a basis set
such as the eigenstates $|x\rangle$ of $\hat{x}$: $|\psi_{i}\rangle=\sum_{x}c_{x}|x\rangle$.

\textbf{Theorem A:} The existence of coherence between outcomes $x_{1}$
and $x_{2}$ of an observable $\hat{x}$ is equivalent, within quantum
mechanics, to the existence of a nonzero off-diagonal element in the
density matrix, i.e, $\left\langle x_{1}\right|\rho\left|x_{2}\right\rangle \neq0$.

\textbf{Proof:} The proof is given in Appendix A. {\scriptsize $\blacksquare$}{\scriptsize \par}

\textbf{Theorem B:} In quantum mechanics, there exists coherence between
outcomes $x_{1}$ and $x_{2}$ of an observable $\hat{x}$ iff in
\emph{any} decomposition (\ref{eq:mixture}) of the density matrix,
there is a nonzero contribution from a superposition state of the
type \begin{equation}
|\psi_{S}\rangle=c_{x_{1}}\left|x_{1}\right\rangle +c_{x_{2}}\left|x_{2}\right\rangle +\sum_{x\neq x_{1},x_{2}}c_{x}\left|x\right\rangle \label{eq:supx}\end{equation}
with $c_{x_{1}}$,$c_{x_{2}}\neq0$. 

\textbf{Proof}: If each $|\psi_{i}\rangle$ cannot be written in the
specific form (\ref{eq:supx}), then each $|\psi_{i}\rangle\langle\psi_{i}|$
is either of form $\rho_{1}$ or $\rho_{2}$, so that we can write
$\rho$ as the mixture (\ref{eq:QM mixture}). Hence the existence
of coherence, which implies $\rho$ cannot be written as (\ref{eq:QM mixture}),
implies the superposition must always exist in (\ref{eq:mixture}).
The converse is also true: if the superposition exists in any decomposition,
then there exists an irreducible term in the decomposition that assigns
nonzero probabilities to both $x_{1}$ and $x_{2}$, and therefore
the density matrix cannot be written as (\ref{eq:QM mixture}). {\scriptsize $\blacksquare$}{\scriptsize \par}

We say that a \emph{generalized $S$-scopic superposition} of states
$|x_{1}\rangle$ and $|x_{2}\rangle$ exists when any decomposition
(\ref{eq:mixture}) must contain a nonzero probability for a superposition
(\ref{eq:supx}), where $x_{1}$ and $x_{2}$ are separated by at
least $S$. Throughout this paper, we define the \emph{size} of the
generalized superposition \begin{equation}
|\psi\rangle=\sum_{k}c_{k}|x_{k}\rangle\label{eq:gmssize}\end{equation}
(where $|x_{k}\rangle$ are eigenstates of $\hat{x}$ and each $c_{k}\neq0$)
to be the range of its prediction for $\hat{x}$, this range being
the maximum value of $|x_{k}-x_{j}|$ where $|x_{k}\rangle$ and $|x_{j}\rangle$
are any two components of the superposition (\ref{eq:gmssize}) (so
$c_{k}$,$c_{j}\neq0$).

From the above discussions it follows that within quantum mechanics,
the existence of generalized $S$-scopic coherence between $x_{1}$
and $x_{2}$ (here $|x_{2}-x_{1}|=S$) implies the existence of a
generalized $S$-scopic superposition of type (\ref{eq:supx}), which
can be written as \begin{equation}
|\psi\rangle=c_{-}\psi_{-}+c_{0}\psi_{0}+c_{+}\psi_{+},\label{eq: QMGMS}\end{equation}
where the quantum state $\psi_{-}$ assigns some nonzero probability
only to outcomes smaller than or equal to $x_{1}$, the quantum state
$\psi_{+}$ assigns some nonzero probability only to outcomes larger
than or equal to $x_{2}$, and the state $\psi_{0}$ assigns nonzero
probabilities only to intermediate values satisfying $x_{1}<x<x_{2}$.
Where $S$ is macroscopic, expression (\ref{eq: QMGMS}) depicts a
\emph{generalized macroscopic superposition} state. In this case then,
only the states $\psi_{-}$ and $\psi_{+}$ are necessarily macroscopically
distinct. We regain the traditional extreme macroscopic quantum state
$c_{-}\psi_{-}+c_{+}\psi_{+}$ when $c_{0}=0$.

\section{Minimum Uncertainty Theories \label{sec:minimum-uncertainty}}

We now follow a procedure similar to that used to derive criteria
useful for the confirmation of inseparability \citep{inseplur}. The
underlying states $\rho_{1}$ and $\rho_{2}$ comprising the mixture
(\ref{eq:QM mixture}) are themselves quantum states, and so each
will satisfy the quantum uncertainty relations with respect to complementary
observables. This and the assumption of Eq. (\ref{eq:QM mixture})
will imply a set of constraints, which take the form of inequalities.
The violation of any one of these is enough to confirm the observation
of a generalized macroscopic coherence---that is, of a generalized
macroscopic superposition of type (\ref{eq: QMGMS}).

While our specific aim is to develop criteria for quantum macroscopic
superpositions, we present the derivations in as general a form as
possible to make the point that experimental violation of the inequalities
would imply not only a generalized macroscopic coherence in quantum
theory, but a failure of the assumption (\ref{eq:probmacro}) in all
theories which place the system in a probabilistic mixture of two
states, which we designate by $\rho_{1}$ and $\rho_{2}$, and for
which the appropriate uncertainty relation holds for each of the states.
In this sense, our approach is similar to that of Bell \citep{bell},
except that the assumption used here of minimum uncertainties for
outcomes of measurements would be regarded as more restrictive than
the local hidden variable theory assumption on which Bell's theorem
is based.

We make this point more specific by defining a whole class of theories,
which we refer to as the minimum uncertainty theories (MUTs), that
embody the assumption that any state $\rho$ within the theory will
predict the same uncertainty relation for the variances of two incompatible
observables $\hat{x}$ and $\hat{p}$ as is predicted by quantum mechanics.
This is a priori not an unreasonable thing to postulate for a theory
that may differ from quantum mechanics in the macroscopic regime but
agree with all the observations in the well-studied microscopic regime.
In this paper we will focus on pairs of observables, like position
and momentum, for which the uncertainty bound is a real number, which
with the use of scaling and choice of units will be set to $1$, so
we can write an uncertainty relation assumed by all MUTs as \begin{equation}
\Delta^{2}x\Delta^{2}p\geq1,\label{eq:MUT UR}\end{equation}
where $\Delta^{2}x$ and $\Delta^{2}p$ are the variances of $x$
and $p$ respectively. This is Heisenberg's uncertainty relation,
and quantum mechanics is clearly a member of MUT. Other quantum uncertainty
relations that will be specifically used in this paper include\begin{equation}
\Delta^{2}x+\Delta^{2}p\geq2,\label{eq:sumhup}\end{equation}
which follows for the same choice of units as that of Eq. (\ref{eq:MUT UR})
and has been useful in derivation of inseparability criteria \citep{inseplur}.

\section{Signatures for generalized S-scopic superpositions: Binned domain}

In this section we will derive inequalities that follow if there are
no $s$-scopic superpositions (where $s>S$), so that violation of
these inequalities implies existence of an $S$-scopic superposition
(or coherence), as defined in Secs. II and III. The approach is similar
to that often used to detect entangled states. Separability implies
inequalities such as those derived by Duan et al. \citet{inseplur},
and their violation thus implies existence of entanglement. This approach
has been used to experimentally confirm entanglement, as described
in Ref. \citet{bowen}, among others. An experimental description
of the approach we use here has been outlined by Marquardt et al.
\citet{marqu}.

We consider two types of criteria for the detection of a generalized
macroscopic superposition (or coherence). The first, of the type considered
in \citep{macsupPRL}, will be considered in this section and uses
\emph{binned outcomes} to demonstrate a generalized $S$-scopic superposition
of states $\psi_{+}$ and $\psi_{-}$ that predict outcomes in \emph{specified}
regions denoted $+1$ and $-1$ respectively (Fig. 2), where these
regions are separated by a minimum distance \emph{$S$.} We expand
on some earlier results of \citep{macsupPRL} for completeness and
also introduce more criteria of this type.

\subsection{Single system}

Consider a system $A$ and a macroscopic measurement $\hat{x}$ on
$A$, the outcomes of which are spread over a macroscopic range. We
partition the domain of outcomes $x$ for this measurement into three
regions, labeled $l=-1,0,1$ for the regions $x\leq-S/2$, $-S/2<x<S/2$,
$x\geq S/2$, respectively. The probabilities for outcomes to fall
in those regions are denoted $\wp_{-},$ $\wp{}_{0}$ and $\wp_{+}$,
respectively (Fig. 2).

\begin{figure}
\includegraphics[scale=0.3]{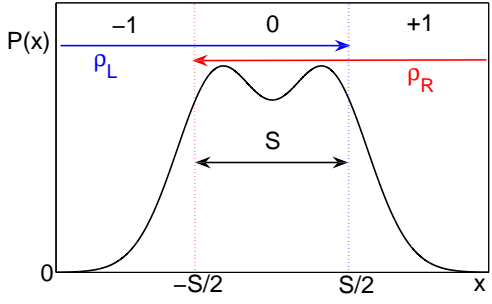}

\caption{Probability distribution for a measurement $\hat{x}$ . We bin results
to give three distinct regions of outcome: $0$, $-1,$$+1$.}

\end{figure}

If there is \emph{no} generalized S-scopic coherence then there is
no coherence between outcomes in $l=1$ and $l=-1$, and the state
of system $A$ can be written as \begin{equation}
\rho_{mix}=\wp_{L}\rho_{L}+\wp_{R}\rho_{R},\label{eq:Binned mixture}\end{equation}
where $\rho_{L}$ predicts outcomes in the region $x<S/2$, $\rho_{R}$
predicts outcomes in the region $x>-S/2$, and $\wp_{L}$ and $\wp_{R}$
are their respective probabilities. The assumption of this mixture
(\ref{eq:Binned mixture}) implies\begin{equation}
P(y)=\wp_{L}P_{L}(y)+\wp_{R}P_{R}(y).\label{eq:probconst}\end{equation}
Here $y$ is the outcome of some measurement that can be performed
on the system, and $P_{R/L}(y)$ is the probability for a result $y$
when the system is specified as being in state $\rho_{R/L}$. Where
the measurement performed is $\hat{x}$, so $y=x$, there is the constraint
on Eq. (\ref{eq:probconst}) so that $P_{R}(x)=0$ for $x\leq-S/2$
and $P_{L}(x)=0$ for $x\geq S/2$.

Now consider an observable $\hat{p}$ (with outcomes $p$) incompatible
with $\hat{x}$, such that the variances are constrained by the uncertainty
relation $\Delta^{2}x\Delta^{2}p\geq1$. Our goal is to derive inequalities
from just two assumptions: firstly, that $\hat{x}$ and $\hat{p}$
are incompatible observables of quantum mechanics (or of a minimum
uncertainty theory), so the uncertainty relation holds for both $\rho_{R/L}$;
and, secondly, that there is no generalized S-scopic coherence.

Violation of these inequalities will imply that one of these assumptions
is false. Within quantum mechanics, for which the first assumption
is necessarily true, that would imply the existence of a generalized
macroscopic superposition of type (\ref{eq: QMGMS}) with outcomes
$x_{1}$ and $x_{2}$ separated by at least $S$.

If the quantum state is of form (\ref{eq:Binned mixture}) or if the
theory satisfies Eq. (\ref{eq:probconst}), then \begin{equation}
\Delta^{2}p\geq\wp_{L}\Delta_{L}^{2}p+\wp_{R}\Delta_{R}^{2}p,\label{eq:mixvarp}\end{equation}
where $\Delta²p$, $\Delta²_{L}p$ and $\Delta²_{R}p$ are the variances
of $p$ in the states $\rho_{mix}$, $\rho_{L}$ and $\rho_{R}$,
respectively. This follows simply from the fact the variance of a
mixture cannot be less than the average variance of its component
states. Specifically, if a probability distribution for a variable
$z$ is of the form $P(z)=\sum_{i=1}^{N}\wp_{i}P_{i}(z)$, then $\Delta^{2}z=\sum_{i=1}^{N}\wp_{i}\Delta_{i}^{2}z+\frac{1}{2}\sum_{i\neq i'}\wp_{i}\wp_{i'}(\langle z\rangle_{i}-\langle z\rangle_{i'})^{2}$. 

We can now, using Eq. (\ref{eq:mixvarp}) and the Cauchy-Schwarz inequality,
derive a bound for a particular function of variances that will apply
if the system is describable as the mixture Eq. (\ref{eq:Binned mixture})\begin{eqnarray}
(\wp_{L}\Delta_{L}^{2}x+\wp_{R}\Delta_{R}^{2}x)\Delta^{2}p & \geq & [\sum_{i=L,R}\wp_{i}\Delta_{i}^{2}x][\sum_{i=L,R}\wp_{i}\Delta_{i}^{2}p]\nonumber \\
 & \geq & [\sum_{i=L,R}\wp_{i}\Delta_{i}x\Delta_{i}p]^{2}\label{eq:cs}\\
 & \geq & 1.\nonumber \end{eqnarray}

The left-hand side is not directly measurable, since it involves variances
of $\hat{x}$ in two states which have overlapping ranges of outcomes.
We must derive an upper bound for $\Delta_{L/R}^{2}x$ in terms of
measurable quantities. For this we partition the probability distribution
$P_{R}(x)$ according to the outcome domains $l=0,1$, into normalized
probability distributions $P_{R0}(x)\equiv P_{R}(x|x<S/2)$ and $P_{+}(x)\equiv P_{R}(x|x\geq S/2)$:\begin{equation}
P_{R}(x)=\wp_{R0}P_{R0}(x)+\wp_{R+}P_{+}(x).\label{eq:domain partition}\end{equation}
Here $\wp_{R+}=\int_{S/2}^{\infty}P_{R}(x)dx=\wp_{+}$ and $\wp_{R0}=\int_{0}^{S/2}P_{R}(x)dx$.
It follows that $\Delta_{R}^{2}x=\wp_{R0}\Delta_{R0}^{2}x+\wp_{R+}\Delta_{+}^{2}x+\wp_{R0}\wp_{R+}(\mu_{+}-\mu_{R0})^{2}$,
where $\mu_{+}$($\Delta_{+}^{2}x$) and $\mu_{R0}$ ($\Delta_{R0}^{2}x$)
are the averages (variances) of $P_{+}(x)$ and $P_{R0}(x)$, respectively.
Using the bounds $\wp_{R0}\leq\wp_{0}/(\wp_{0}+\wp_{+})$, $\Delta_{R0}^{2}x\leq S^{2}/4$,
$\wp_{R+}\leq1$ and $0\leq\mu_{+}-\mu_{R0}\leq\mu_{+}+S/2$, we derive
\begin{equation}
\Delta_{R}^{2}x\leq\Delta_{+}^{2}x+\frac{\wp_{0}}{\wp_{0}+\wp_{+}}[(S/2)^{2}+(\mu_{+}+S/2)^{2}]\label{eq:domainbound}\end{equation}
and, by similar reasoning,\begin{equation}
\Delta_{L}^{2}x\leq\Delta_{-}^{2}x+\frac{\wp_{0}}{\wp_{0}+\wp_{-}}[(S/2)^{2}+(\mu_{-}-S/2)^{2}].\label{eq:domainbound-}\end{equation}
 Here $\mu_{\pm}$ and $\Delta_{\pm}^{2}x$ are the mean and variance
of the measurable $P_{\pm}(x)$, which, since the only contributions
to the regions + and - are from $P_{R}(x)$ and $P_{L}(x)$ respectively,
are equal to the normalized $+$ and $-$ parts of $P(x)$, so that
$P_{+}(x)=P(x|x\geq S/2)$ and $P_{-}(x)=P(x|x\leq-S/2)$. We substitute
Eq. (\ref{eq:domainbound}) in Eq. (\ref{eq:cs}), and use $\wp_{0}+\wp_{+}\geq\wp_{R}$
and $\wp_{0}+\wp_{-}\geq\wp_{L}$ to derive the final result which
is expressed in the following theorem.

\textbf{Theorem 1}: The assumption of no generalized $S$-scopic coherence
between outcomes in regions $+1$ and $-1$ of Fig. 2 (or, equivalently,
of no generalized $S$-scopic superpositions involving two states
$\psi_{-}$ and $\psi_{+}$ predicting outcomes for $\hat{x}$ in
the respective regions $+1$ and $-1$) will imply the uncertainty
relations\begin{equation}
(\Delta_{ave}^{2}x+\wp_{0}\delta)\Delta^{2}p\geq1\label{eq:critsuper}\end{equation}
and\begin{equation}
\Delta_{ave}^{2}x+\Delta^{2}p\geq2-\wp_{0}\delta,\label{eq:sumcrit}\end{equation}
where we define $\Delta_{ave}^{2}x=\wp_{+}\Delta_{+}^{2}x+\wp_{-}\Delta_{-}^{2}x$
and $\delta\equiv\{(\mu_{+}+S/2)^{2}+(\mu_{-}-S/2)^{2}+S²/2\}+\Delta_{+}^{2}x+\Delta_{-}^{2}x$.
Thus, the violation of either one of these inequalities implies the
existence of a generalized $S$-scopic quantum superposition, and
in this case the superposition involves states $\psi_{+}$ and $\psi_{-}$
predicting outcomes for $\hat{x}$ in regions $+1$ and $-1$, of
Fig. 2, respectively.

As illustrated in Fig.2, the $\Delta_{\pm}^{2}x$ and $\mu_{\pm}$
are the variance and mean of $P_{\pm}(x)$, the normalized distribution
over the domain $l=\pm1$. $\wp_{\pm}$ is the total probability for
a result $x$ in the domain $l=\pm1$, while $\wp_{0}=1-(\wp_{+}+\wp_{-})$.
The measurement of the probability distributions for $\hat{x}$ and
$\hat{p}$ are all that is required to determine whether violation
of the inequality (\ref{eq:critsuper}) or (\ref{eq:sumcrit}) occurs.
Where $\hat{x}$ and $\hat{p}$ correspond to optical field quadratures,
such distributions have been measured, for example, by Smithey \emph{et
al.} \citep{gaus}.

\textbf{Proof}: The assumption of no such generalized $S$-scopic
superposition implies Eq. (\ref{eq:Binned mixture}). We have proved
that Eq. (\ref{eq:critsuper}) follows. To prove Eq. (\ref{eq:sumcrit}),
we start from Eq. (\ref{eq:Binned mixture}) and the uncertainty relation
(\ref{eq:sumhup}), and derive a bound that will apply if the system
is describable as Eq. (\ref{eq:Binned mixture}): $(\wp_{L}\Delta_{L}^{2}x+\wp_{R}\Delta_{R}^{2}x)+\Delta^{2}p\geq[\sum_{i=L,R}\wp_{i}\Delta_{i}^{2}x]+[\sum_{i=L,R}\wp_{i}\Delta_{i}^{2}p]\geq[\sum_{i=L,R}\wp_{i}[\Delta_{i}^{2}x+\Delta_{i}^{2}p]\geq2$.
Using (\ref{eq:domainbound}), (\ref{eq:domainbound-}) and $\wp_{0}+\wp_{+}\geq\wp_{R}$
and $\wp_{0}+\wp_{-}\geq\wp_{L}$ we get the final result. {\scriptsize $\blacksquare$}{\scriptsize \par}

\subsection{Bipartite systems}

One can derive similar criteria where we have a system comprised of
two subsystems $A$ and $B$. In this case, a reduced variance may
be found in a combination of observables from both subsystems. A common
example is where there is a correlation between the two positions
$X^{A}$ and $X^{B}$ of subsystems $A$ and $B$ respectively, and
also between the two momenta $P^{A}$ and $P^{B}$. Such correlation
was discussed by Einstein, Podolsky and Rosen \citep{epr} and is
called EPR correlation. If a sufficiently strong correlation exists,
it is possible that both the position difference $X^{A}-X^{B}$ and
the momenta sum $P^{A}+P^{B}$ will have zero variance.

Where we have two subsystems that may demonstrate EPR correlation,
we may construct a number of useful complementary measurements that
may reveal generalized macroscopic superpositions. The simplest situation
is where we again consider superpositions with respect to the observable
$X^{A}$ of system $A$. Complementary observables include observables
of the type \begin{equation}
\tilde{P}=P^{A}-gP^{B},\label{eq:ptilde}\end{equation}
 where $g$ is an arbitrary constant and $P^{B}$ is an observable
of system $B$. We denote the outcomes of measurements $X^{A},$ $P^{A},$
$P^{B},$ $\tilde{P}$ by the lower case symbols $x^{A},$ $p^{A},$
$p^{B},$ $\tilde{p}$ respectively. The Heisenberg uncertainty relation
is\begin{equation}
\Delta^{2}x^{A}\Delta_{inf,L}^{2}p^{A}=\Delta^{2}x^{A}\Delta^{2}\tilde{p}\geq1.\label{eq:uncerinf}\end{equation}
We have introduced $\Delta_{inf,L}^{2}p^{A}=\Delta^{2}\tilde{p}$
so that a connection is made with notation used previously in the
context of demonstration of the EPR paradox \citep{eprr,eprreview}.
More generally \citep{mdrepr,eprreview}, we define an inference variance
\begin{equation}
\Delta_{inf}^{2}p^{A}=\sum_{p^{B}}P(p^{B})\Delta^{2}(p^{A}|p^{B}),\label{eq:condvar}\end{equation}
which is the average conditional variance for $P^{A}$ at $A$ given
a measurement of $P^{B}$ at $B$. The $\Delta^{2}(p^{A}|p^{B})$
are the variances of the conditional probability distributions $P(p^{A}|p^{B}).$
We note that $\Delta_{inf,L}^{2}p^{A}$ is the linear regression estimate
of $\Delta_{inf}^{2}p^{A}$, but that we have $\Delta_{inf}^{2}p^{A}=\Delta_{inf,L}^{2}p^{A}$
for the case of Gaussian states \citep{eprreview}. The uncertainty
relation \begin{equation}
\Delta^{2}x^{A}\Delta_{inf}^{2}p^{A}\geq1\label{eq:condinfur}\end{equation}
and also $\Delta^{2}p^{A}\Delta_{inf}^{2}x^{A}\geq1$, holds true
for all quantum states \citep{newjournmidopt}, so that we can interchange
$\Delta_{inf}^{2}p^{A}$ with $\Delta_{inf,L}^{2}p^{A}$ in the proofs
and theorems below.

\textbf{Theorem 2}: Where we have a system comprised of subsystems
$A$ and $B$, the absence of generalized $S$-scopic superpositions
with respect to the measurement $X^{A}$ implies\begin{equation}
(\Delta_{ave}^{2}x^{A}+\wp_{0}\delta)\Delta_{inf}^{2}p^{A}\geq1.\label{eqn:eprcat}\end{equation}
$\Delta_{ave}^{2}x^{A}$, $\wp_{0}$ and $\delta$ are defined as
for theorem 1 for the distribution $P(x^{A})$. $\Delta_{inf}^{2}p^{A}$
is defined by Eq. (\ref{eq:condvar}) and involves measurements performed
on both systems $A$ and $B$. The inequality Eq. (\ref{eqn:eprcat})
also holds replacing $\Delta_{inf}^{2}p^{A}$ with $\Delta_{inf,L}^{2}p^{A}$
which is defined by Eq. (\ref{eq:uncerinf}). Thus violation of Eq.
(\ref{eqn:eprcat}) implies the existence of the generalized $S$-scopic
superposition, involving states predicting outcomes for $X^{A}$ in
regions $+1$ and $-1$.

\textbf{Proof}: The proof follows in identical fashion to that of
theorem 1, except in this case the $\rho_{L}$ and $\rho_{R}$ of
Eq. (\ref{eq:Binned mixture}) are states of the composite system,
and there is no constraint on these except that the domain for outcomes
of $X^{A}$ is restricted as specified in the definition of $\rho_{R/L}$.
The expansion (\ref{eq:mixture}) for the density matrix as a mixture
is $\rho=\sum_{r}\wp_{r}|\psi_{r}\rangle\langle\psi_{r}|$ where now
$\psi_{r}=\sum_{i,j}c_{i,j}|x_{i}\rangle_{A}|x_{j}\rangle_{B},$ $|x_{j}\rangle_{B}$
being eigenstates of an observable of system $B$ that form a basis
set for states of $B$. The generalized superposition (\ref{eq:supx})
thus becomes in this bipartite case\begin{equation}
|\psi_{r}\rangle=c_{1}\left|x_{1}\right\rangle _{A}|u_{1}\rangle_{B}+c_{2}\left|x_{2}\right\rangle _{A}|u_{2}\rangle_{B}+\sum_{i\neq1,2}c_{ij}\left|x_{i}\right\rangle _{A}|x_{j}\rangle_{B},\label{eq:supbi}\end{equation}
where $|u_{1}\rangle$ and $|u_{2}\rangle$ are pure states for system
$B$. If we assume no generalized $S$-scopic superposition, then
$\rho$ can be written without contribution from a state of form (\ref{eq:supbi})
and we can write $\rho$ as Eq. (\ref{eq:Binned mixture}). The constraint
(\ref{eq:Binned mixture}) implies $P(\tilde{p})=\sum_{I=R,L}\wp_{I}P_{I}(\tilde{p})$
where $P_{R|L}(\tilde{p})$ is the probability distribution of $\tilde{p}$
for state $\rho_{R/L}$. Thus Eq. (\ref{eq:mixvarp}) also holds for
$\tilde{p}$ replacing $p$, as do all the results (\ref{eq:domain partition})-(\ref{eq:domainbound-})
involving the variances of $x^{A}$. Also, Eq. (\ref{eq:mixvarp})
holds for $\Delta_{inf}^{2}p^{A}$ (see Appendix B). Thus we prove
theorem 2 by following Eqs. (\ref{eq:mixvarp})-(\ref{eq:critsuper}).
{\scriptsize $\blacksquare$}{\scriptsize \par}

In order to violate the inequality (\ref{eqn:eprcat}), we would look
to minimize $\Delta_{inf}^{2}p^{A}$, or $\Delta_{inf,L}^{2}p^{A}=\Delta^{2}\tilde{p}$.
For the optimal EPR states, $P^{A}+P^{B}$ has zero variance, and
one would choose for $\tilde{P}$ the case of $g=-1$, so that $\tilde{p}=p^{A}+p^{B}$,
where $p^{B}$ is the result of measurement of $P^{B}$ at $B$. This
case gives $\Delta_{inf}^{2}p^{A}=0$. More generally for quantum
states that are not the ideal case of EPR, our choice of $\tilde{p}$
becomes so as to optimize the violation of Eq. (\ref{eqn:eprcat})
and will depend on the quantum state considered. This will be explained
further in Sec. VIII.

A second approach is to use as the macroscopic measurement a linear
combination of observables from both systems $A$ and $B$, so for
example we might have $\hat{x}=(X^{A}+X^{B})/\sqrt{2}$ and $\hat{p}=(P^{A}+P^{B})/\sqrt{2}$.
Relevant uncertainty relations include (based on $|[X^{A},P^{A}]|=2$
which gives $\Delta x^{A}\Delta p^{A}=1$) \begin{equation}
\Delta(x^{A}+x^{B})\Delta(p^{A}+p^{B})\geq2\label{eq:prodsum}\end{equation}
and \begin{equation}
\Delta^{2}(x^{A}+x^{B})+\Delta^{2}(p^{A}+p^{B})\geq4.\label{eq:sumsumur}\end{equation}
and from these we can derive criteria for generalized S-scopic coherence
and superpositions. 

\textbf{Theorem 3}: The following inequalities if violated will imply
existence of generalized S-scopic superpositions \begin{equation}
\Bigl(\Delta_{ave}^{2}(\frac{x^{A}+x^{B}}{\sqrt{2}})+\wp_{0}\delta\Bigr)\Delta^{2}(\frac{p^{A}+p^{B}}{\sqrt{2}})\geq1\label{eq:critsuper bipartite}\end{equation}

and\begin{equation}
\Delta_{ave}^{2}(\frac{x^{A}+x^{B}}{\sqrt{2}})+\Delta^{2}(\frac{p^{A}+p^{B}}{\sqrt{2}})\geq2-\wp_{0}\delta.\label{eq:sumcrit bipartite}\end{equation}
We write in terms of the normalized quadratures so that, following
Eq. (\ref{eq:prodsum}), $\Delta^{2}(\frac{x^{A}+x^{B}}{\sqrt{2}})<1$
would imply squeezing of the variance below the quantum noise level.
The quantities $\Delta_{ave}^{2}x$, $\wp_{0}$ and $\delta$ are
defined as for theorem 1, but we note that $P(x)$ in this case is
the distribution for $\hat{x}=(X^{A}+X^{B})/\sqrt{2}$. $S$ now refers
to the size of the superposition of $(X^{A}+X^{B})/\sqrt{2}$. 

\textbf{Proof:} In this case the $\rho_{R/L}$ of Eq. (\ref{eq:Binned mixture})
are defined as specified originally in (\ref{eq:Binned mixture})
but where $x$ is now defined as the outcome of the measurement $\hat{x}=(X^{A}+X^{B})/\sqrt{2}$.
The failure of the form (\ref{eq:Binned mixture}) for $\rho$ is
equivalent to the existence of a generalized superposition of type
(\ref{eq:supbi}) where now $|x_{i}\rangle$ refers to eigenstates
of $X^{A}+X^{B}$. Thus the eigenstates $|x_{i}\rangle$ are of the
general form $|x_{i}\rangle=\sum_{x_{j}}c_{j}|x_{j}\rangle_{A}|x_{i}-x_{j}\rangle_{B}$.
The mixture (\ref{eq:Binned mixture}) implies Eq. (\ref{eq:mixvarp})
where now $p$ refers to the outcome of $\hat{p}=(P^{A}+P^{B})\sqrt{2}$,
and will imply a similar inequality for $\hat{x}$. Application of
uncertainty relation (\ref{eq:prodsum}) for the products can be used
in Eq. (\ref{eq:cs}), and the proof of the theorem follows as in
(\ref{eq:mixvarp})-(\ref{eq:critsuper}) of theorem 1. The second
result follows by applying the procedure for proof of Eq. (\ref{eq:sumcrit})
but using the sum uncertainty relation (\ref{eq:sumsumur}). {\scriptsize $\blacksquare$}{\scriptsize \par}

\section{Signatures of non-locatable generalized S-scopic superpositions}

A second set of criteria will be developed, to demonstrate that a
generalized S-scopic superposition exists, so that two states comprising
the superposition predict respective outcomes separated by at least
size S, but in this case there is the disadvantage that no information
is obtained regarding the regions in which these outcomes lie. 

This lack of information is compensated by a far simpler form of the
inequalities and increased sensitivity of the criteria. For pure states,
a measurement of squeezing $\Delta p$ implies a state that when written
in terms of the eigenstates of $x$ is a superposition such that $\Delta x\geq1/\Delta p$.
With increasing squeezing, the extent $S$ of the superposition increases.
To develop a simple relationship between $S$ and $\Delta p$ for
mixtures, we assume that there is no such generalized coherence between
any outcomes of $\hat{x}$ separated by a distance larger than $S$.
This approach gives a simple connection between the minimum size of
a superposition describing the system and the degree of squeezing
that is measured for this system. The drawback is the loss of direct
information about the location (in phase space for example) of the
superposition. We thus refer to these superpositions as \char`\"{}non-locatable\char`\"{}.

\subsection{Single systems}

We consider the outcome domain of a macroscopic observable $\hat{x}$
as illustrated in Fig. 3, and address the question of whether this
distribution could be predicted from microscopic, or $s$-scopic ($s<S$),
superpositions of eigenstates of $\hat{x}$ alone. 

\begin{figure}
\includegraphics[scale=0.3]{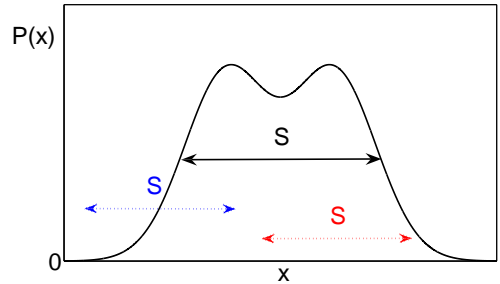}

\caption{We consider an arbitrary probability distribution for a measurement
$\hat{x}$ that gives a macroscopic range of outcomes. }

\end{figure}

The assumption of no generalized $S$-scopic coherence (between any
two outcomes of the domain for $\hat{x}$) or, equivalently, the assumption
of no generalized $S$-scopic superpositions, with respect to eigenstates
of $\hat{x}$, means that the state can be written in the form\begin{equation}
\rho_{S}=\sum_{i}\wp_{i}\rho_{Si}.\label{eq:mixmicrosup}\end{equation}
Here each $\rho_{Si}$ is the density operator for a pure quantum
state that is \emph{not} such a generalized $S$-scopic superposition,
so that $\rho_{Si}$ has a range of possible outcomes for $\hat{x}$
separated by less than $S$. Hence $\rho_{Si}=|\psi_{Si}\rangle\langle\psi_{Si}|$
where \begin{equation}
|\psi_{Si}\rangle=\sum_{k}c_{k}|x_{k}\rangle\label{eq:microsup}\end{equation}
but the maximum separation of any two states $|x_{k}\rangle$,$|x_{k'}\rangle$
involved in the superposition (that is with $c_{k},c_{k'}\neq0$ )
is less than $S$, so $|x_{k}-x_{k'}|<S$. 

Assumption (\ref{eq:mixmicrosup}) will imply a constraint on the
measurable statistics, namely that there is a minimum level of uncertainty
in the prediction for the complementary observable $\hat{p}$. The
variances of each $\rho_{Si}$ must be bounded by \begin{equation}
\Delta²_{Si}x<\frac{S²}{4}.\label{eq:dommicrosup}\end{equation}
It is also true that \begin{equation}
\Delta^{2}p\geq\sum_{i}\wp_{i}\Delta²_{Si}p.\label{eq:mixmomentum}\end{equation}
Now the Heisenberg uncertainty relation applies to each $\rho_{Si}$
(the inequality also applies to the MUT's discussed in Sec. \ref{sec:minimum-uncertainty})
so for the incompatible observables $\hat{x}$ and $\hat{p}$\begin{equation}
\Delta²_{Si}x\Delta²_{Si}p\geq1.\label{eq:hupfor proofgen}\end{equation}
Thus a lower bound on the variance of $p$ follows: \begin{eqnarray}
\Delta^{2}p & \geq & \sum_{i}\wp_{i}\Delta²_{Si}p\label{eq:minfuzp}\\
 & \geq & \sum_{i}\wp_{i}\frac{1}{\Delta_{Si}^{2}x}>\frac{4}{S^{2}}.\nonumber \end{eqnarray}
We thus arrive at the following theorem.

\textbf{Theorem 4:} The assumption of no generalized $S$-scopic coherence
in $\hat{x}$ will imply the following inequality for the variance
of outcomes of the complementary observable $\hat{p}$\begin{equation}
\Delta p>\frac{2}{S}.\label{eq:nobinning ineq}\end{equation}
The main result of this section follows from theorem 4 and is that
the observation of a squeezing $\Delta p$ in $\hat{p}$ such that
\begin{equation}
\Delta p\leq2/S\label{eq:critsqsup}\end{equation}
will imply the existence of an $S$- scopic superposition\begin{equation}
c_{x}|x\rangle+c_{x+S}|x+S\rangle+......\label{eq:supS}\end{equation}
namely, of a superposition of eigenstates $|x\rangle$ of $\hat{x}$,
that give predictions for $\hat{x}$ with a range of at least $S$.
The parameter $S$ gives a minimum extent of quantum indeterminacy
with respect to the observable $\hat{x}$. Here $c_{x}$ and $c_{x+S}$
represent non-zero probability amplitudes.

In fact, using our criterion (\ref{eq:critsqsup}) squeezing in $p$
($\Delta p<1$) will rule out any expansion of the system density
operator in terms of superpositions of $|x\rangle$ with $S\leq2$
(Fig. 4). Thus onset of squeezing is evidence of the onset of quantum
superpositions of size $S>2$, the size $S=2$ corresponding to the
vacuum noise level. This noise level may be taken as a level of reference
in determining the relative size of the superposition. The experimental
observation \citep{squresultexp} of squeezing levels of $\Delta p\approx0.4$
confirms superpositions of size at least $S=5$.

\subsection{Bipartite systems}

For composite systems comprised of two subsystems $A$ and $B$ upon
which measurements $X^{A}$, $P^{A}$, $X^{B}$, $P^{B}$ can be performed,
the approach of the previous section leads to the following theorems.

\textbf{Theorem 5a}. The assumption of no generalized $S$-scopic
coherence with respect to $X^{A}$ implies \begin{equation}
\Delta_{inf}p^{A}>\frac{2}{S}.\label{eq:nobinning ineq2}\end{equation}
$\Delta_{inf}^{2}p^{A}$ is defined as in Eq. (\ref{eq:condvar}).
The result also holds on replacing $\Delta_{inf}^{2}p$ with $\Delta_{inf,L}^{2}p$
as defined in Eq. (\ref{eq:uncerinf}).

\textbf{Theorem 5b}. The assumption of no generalized $S$-scopic
coherence with respect to $\hat{x}=(X^{A}+X^{B})/\sqrt{2}$ implies\begin{equation}
\Delta(\frac{p^{A}+p^{B}}{\sqrt{2}})>\frac{2}{S}.\label{eq:resultcompsum}\end{equation}

\textbf{Proof:} The proofs follow as for theorem 4, but using the
uncertainty relations (\ref{eq:uncerinf}) and (\ref{eq:prodsum})
in Eq. (\ref{eq:minfuzp}) instead of Eq. (\ref{eq:hupfor proofgen}).
{\scriptsize $\blacksquare$}{\scriptsize \par}

The observation of squeezing such that Eq. \eqref{eq:nobinning ineq2}
is violated will imply the existence of an $S$-scopic superposition\begin{equation}
c_{x}|x\rangle_{A}|u_{1}\rangle_{B}+c_{x+S}|x+S\rangle_{A}|u_{2}\rangle_{B}+......\label{eq:supSent}\end{equation}
namely, of a superposition of eigenstates $|x\rangle_{A}$ that give
predictions for $X^{A}$ separated by at least $S$. Similarly, the
observation of two-mode squeezing such that Eq. \eqref{eq:resultcompsum}
is violated will imply existence of an $S$-scopic superposition of
eigenstates of the normalized position sum $(X^{A}+X^{B})/\sqrt{2}$.

\section{Criteria for generalized $S$-scopic coherent state superpositions }

The criteria developed in the previous section may be used to rule
out that a system is describable as a mixture of coherent states,
or certain superpositions of them. If a system can be represented
as a mixture of coherent states $|\alpha\rangle$ the density operator
for the quantum state will be expressible as\begin{equation}
\rho=\int P(\alpha)|\alpha\rangle\langle\alpha|d{}^{2}\alpha\label{eq:GSrep}\end{equation}
which is, since $P(\alpha)$ is positive for a mixture, the Glauber-Sudarshan
P-representation \citep{Glauber-Sudarshan}. The quadratures $\hat{x}$
and $\hat{p}$ are defined as $x=a+a^{\dagger}$ and $p=(a-a^{\dagger})/i$,
so that $\Delta x=\Delta p=1$ for this minimum uncertainty state,
where here $a$, $a^{\dagger}$ are the standard boson creation and
annihilation operators, so that $a|\alpha\rangle=\alpha|\alpha\rangle$.
Proving failure of mixtures of these coherent states would be a first
requirement in a search for macroscopic superpositions, since such
mixtures expand the system density operator in terms of states with
equal yet minimum uncertainty in each of $x$ and $p$, that therefore
do not allow significant macroscopic superpositions in either.

The coherent states form a basis for the Hilbert space of such bosonic
fields, and any quantum density operator can thus be expanded as a
mixture of coherent states or their superpositions. It is known \citep{dannature}
that systems exhibiting squeezing ($\Delta p<1$) cannot be represented
by the Glauber-Sudarshan representation, and hence onset of squeezing
implies the existence of some superposition of coherent states. A
next step is to rule out mixtures of \emph{$s_{\alpha}$-scopic superpositions}
of coherent states . To define what we mean by this, we consider superpositions
\begin{equation}
|\psi_{s_{\alpha}}\rangle=\sum_{i}c_{i}|\alpha_{i}\rangle\label{eq:cohssup}\end{equation}
where for any $|\alpha_{i}\rangle$, $|\alpha_{j}\rangle$ such that
$c_{i},c_{j}\neq0$, we have $|\alpha_{i}-\alpha_{j}|\leq s_{\alpha}$
for all $i$, $j$ ($s_{\alpha}$ is a positive number). We note that
for a coherent state $|\alpha\rangle,$ $\langle x\rangle=2\alpha$.
Thus the separation of the states with respect to $\hat{x}$ is defined
as $S_{\alpha}=2s_{\alpha}$. The {}``separation'' of the two coherent
states $|-\alpha\rangle$ and $|\alpha\rangle$ (where $\alpha$ is
real) in terms of $x$ corresponds to $S_{\alpha}=4\alpha=2s_{\alpha}$,
as illustrated in Fig. 5. 

\begin{figure}
\includegraphics[scale=0.3]{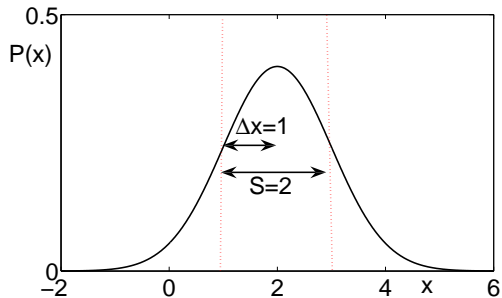}

\caption{$P(x)$ for a coherent state $|\alpha\rangle$: $\Delta x=\Delta p=1$. }

\end{figure}

We next ask whether the density operator for the system can be described
in terms of the $s_{\alpha}$- scopic coherent superpositions, so
that \begin{equation}
\rho=\sum_{r}\wp_{r}|\psi_{s_{\alpha}}^{r}\rangle\langle\psi_{s_{\alpha}}^{r}|\label{eq:mixcohS}\end{equation}
where each $|\psi_{s_{\alpha}}^{r}\rangle$ is of the form (\ref{eq:cohssup}).
Each $|\psi_{s_{\alpha}}^{r}\rangle$ predicts a variance in $x$
which has an upper limit given by that of the superposition $(1/\sqrt{2})\{e^{i\pi/4}|-s_{\alpha}/2\rangle+e^{-i\pi/4}|s_{\alpha}/2\rangle\}$.
This state predicts a probability distribution $P(x)=\frac{1}{2}\sum_{\pm}P_{G\pm}(x)$
where \begin{equation}
P_{G\pm}(x)=\frac{1}{\sqrt{2\pi}}\exp[\frac{-(x\mp s_{\alpha})^{2}}{2}]\label{eq:coherent-state prob}\end{equation}
(Fig.5), which corresponds to a variance $\Delta^{2}x=\langle x^{2}\rangle=1+s_{\alpha}^{2}=1+S_{\alpha}^{2}/4$.
This means each $|\psi_{s_{\alpha}}^{r}\rangle$ is constrained to
allow only $\Delta^{2}x\leq1+s_{\alpha}^{2}$, which implies for
each $|\psi_{s_{\alpha}}^{r}\rangle$ a lower bound on the variance
$\Delta^{2}p$ so that $\Delta^{2}p\geq1/\Delta^{2}x\geq1/(1+s_{\alpha}^{2})$.
Thus using the result for a mixture (\ref{eq:mixcohS}), we get that
if indeed Eq. (\ref{eq:mixcohS}) can describe the system, the variance
in $p$ is constrained to satisfy $\Delta^{2}p\geq1/(1+s_{\alpha}^{2})$. 

Thus observation of squeezing $\Delta^{2}p<1$, so that the inequality\begin{equation}
\Delta^{2}p<1/(1+s_{\alpha}^{2})\label{eq:supcohresult}\end{equation}
is violated, will allow deduction of superpositions of coherent states
with separation at least $s_{\alpha}$. This separation corresponds
to a separation of $S_{\alpha}=2s_{\alpha}$ in $x$ between the two
corresponding Gaussian distributions (Fig. 5), on the scale where
$\Delta^{2}x=1$ is the variance predicted by each coherent state. 

We note that measured values of squeezing $\Delta p\approx0.4$ \citep{squresultexp}
would imply $s_{\alpha}\gtrsim2.2$. This confirms the existence of
a superposition of type \begin{equation}
|\psi_{S}\rangle=\sum_{i}c_{i}|\alpha_{i}\rangle=c_{-}|-\alpha_{0}\rangle+...+c_{+}|\alpha_{0}\rangle\label{eq:gencohsupstexp}\end{equation}
where a separation of at least $s_{\alpha}=|\alpha_{i}-\alpha_{j}|=2.2$
occurs between two coherent states comprising the superposition, so
that we may write $\alpha_{0}=1.1$. Note we have defined reference
axes in phase space selected so that the $x$ axis is the line connecting
the two most separated states $|\alpha_{i}\rangle$ and $|\alpha_{j}\rangle$
so that $|\alpha_{i}-\alpha_{j}|=2\alpha_{0}$ and the $p$ axis cuts
bisects this line. Equation (\ref{eq:gencohsupstexp}) can be compared
with experimental reports \citep{philcats} of generation of extreme
coherent superpositions of type $(1/\sqrt{2})\{e^{i\pi/4}|-\alpha_{0}\rangle+e^{-i\pi/4}|\alpha_{0}\rangle\}$
where $|\alpha_{0}|^{2}=0.79$, implying $\alpha_{0}=0.89$. The corresponding
generalized $s_{\alpha}-$scopic superposition (\ref{eq:gencohsupstexp})
as confirmed by the squeezing measurement involves at least the two
extreme states with $|\alpha_{0}|^{2}=1.2$, but could include other
coherent states with $|\alpha_{0}|<1.1$.

\begin{figure}
\includegraphics[scale=0.3]{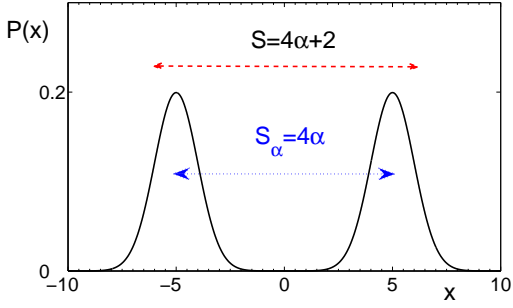}

\caption{(a) $P(x)$ for a superposition of coherent states $(1/\sqrt{2})\{e^{i\pi/4}|-\alpha\rangle+e^{-i\pi/4}|\alpha\rangle\}$
(here the scale is such that $\Delta x=1$ for the coherent state
$|\alpha\rangle$). }

\end{figure}

\section{predictions of particular quantum states}

We will now consider experimental tests of the inequalities derived
above. An important point is that the criteria presented are \emph{sufficient}
to prove the existence of generalized macroscopic superpositions,
but there are many macroscopic superpositions which do not satisfy
the above criteria. Nevertheless there are some systems of current
experimental interest which do allow for violation of the inequalities.
We analyse such cases below, noting that the violation would be predicted
without the experimenter needing to make assumptions about the particular
state involved.

\subsection{Coherent states}

The wave function for the coherent state $|\alpha\rangle$ is \begin{equation}
\left\langle x|\alpha\right\rangle =\frac{1}{(2\pi)^{\frac{1}{4}}}\exp\{\frac{-x^{2}}{4}+\alpha x-|\alpha|^{2}\}.\label{eq:cohwfun}\end{equation}
This gives the expansion in the continuous basis set $|x\rangle$,
the eigenstates of $\hat{x}$. Thus for the coherent state\begin{equation}
|\alpha\rangle=\sum_{x}c_{x}|x\rangle=\int\langle x|\alpha\rangle|x\rangle dx\label{eq:cohexp}\end{equation}
The probability distribution for $x$ is the Gaussian (Fig. 4)\begin{equation}
P(x)=|\langle x|\alpha\rangle|^{2}=\frac{1}{(2\pi)^{\frac{1}{2}}}\exp\{\frac{-(x-2\alpha)^{2}}{2}\}\label{eq:cohgauspx}\end{equation}
(we take $\alpha$ to be real) centered at $2\alpha$ and with variance
$\Delta^{2}x=1.$

\begin{figure}
\includegraphics[scale=0.3]{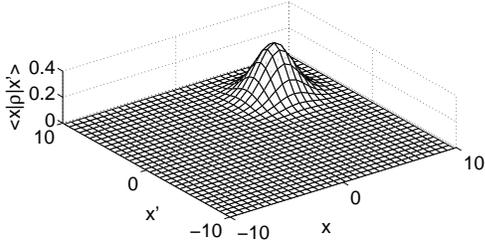}

\caption{Plot of $\langle x|\rho|x'\rangle$ for a coherent state $|\alpha\rangle$,
where $\alpha=2.5$.}

\end{figure}

The coherent state possesses nonzero off-diagonal elements $\langle x|\rho|x'\rangle$
where $|x-x'|$ is large and thus strictly speaking can be regarded
as a generalized macroscopic superposition. However, as $x$ and $x'$
deviate from $2\alpha$, the matrix elements decay rapidly, and the
off-diagonal elements decay rapidly with increasing separation. \begin{equation}
\langle x|\rho|x'\rangle=\frac{1}{(2\pi)^{\frac{1}{2}}}\exp\{\frac{-(x-2\alpha)^{2}}{4}+\frac{-(x'-2\alpha)^{2}}{4}\}\label{eq:cohdiaele}\end{equation}
In effect then, the off-diagonal elements become zero for significant
separations $|x-x'|\geq1$ (Fig.6). We can expect that the detection
of the macroscopic aspects of this superposition will be difficult.
Since $\Delta p=1$, it follows that we can use the criterion (\ref{eq:nobinning ineq})
to prove coherence between outcomes of $x$ separated by at most $S=2$
(Fig. 4), which corresponds to the separation $S=2\Delta x$.

\subsection{Superpositions of coherent states}

The superposition of two coherent states \citep{yurkestol}\begin{equation}
|\psi\rangle=(1/\sqrt{2})\{e^{i\pi/4}|-\alpha\rangle+e^{-i\pi/4}|\alpha\rangle\}\label{eq:scatys}\end{equation}
where $\alpha$ is real and large is an example of a macroscopic superposition
state. The wave function in the position basis is \[
\langle x|\psi\rangle=\frac{-ie^{i\pi/4}e^{[-x^{2}/4-\alpha^{2}]}}{\sqrt{2}(2\pi)^{\frac{1}{4}}}\{e^{\alpha x}+ie^{-\alpha x}\}\]

We consider the two complementary observables $\hat{x}$ and $\hat{p}$,
and note that the probability distribution $P(x)$ for $\hat{x}$
displays two Gaussian peaks centered on $x=\pm2\alpha$ (Fig.5): $P(x)=\frac{1}{2}\sum_{\pm}P_{G\pm}(x)$
where $P_{G\pm}(x)=\exp[{-(x\mp2\alpha)^{2}/2}]/\sqrt{2\pi}$. Each
Gaussian has variance $\Delta^{2}x=1$.

\begin{figure}
\includegraphics[scale=0.3]{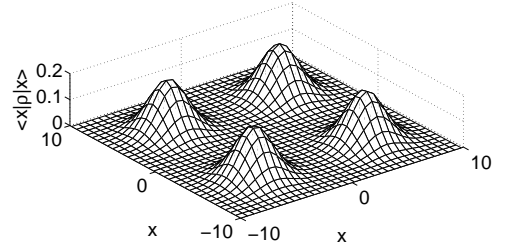}

\caption{Plot of $\langle x|\rho|x'\rangle$ for the superposition state (\ref{eq:scatys}),
where $\alpha=2.5$. }

\end{figure}

The macroscopic nature of the superposition is reflected in the significant
magnitude of the off-diagonal elements $\langle x|\rho|x'\rangle$
where $x=\pm2\alpha$ and $x'=\mp2\alpha$, corresponding to $|x-x'|=4\alpha$.
In fact \begin{equation}
|\langle x|\rho|x'\rangle|=\frac{e^{\frac{-(x^{2}+x'^{2})}{4}-2\alpha^{2}}}{\sqrt{2\pi}}\sqrt{\cosh(2\alpha x)\cosh(2\alpha x')}\label{eq:supcohdiaele}\end{equation}
 as plotted in Fig. 7 and which for these values of $x$ and $x'$
becomes $\frac{(1-e^{-8\alpha^{2}})}{2(2\pi)^{\frac{1}{2}}}$. With
significant off-diagonal elements connecting macroscopically different
values of $x$, this superposition is a good example of a generalized
macroscopic superposition (\ref{eq: QMGMS}). 

Nonetheless we show that the simple linear criteria (\ref{eq:nobinning ineq})
and (\ref{eq:critsuper}) derived from Eq. (\ref{eq:mixture}) are
not sufficiently sensitive to detect the extent of the macroscopic
coherence of this superposition state (\ref{eq:scatys}), even though
the state (\ref{eq:scatys}) cannot be written in the form (\ref{eq:Binned mixture}).
We point out that it may be possible to derive further nonlinear constraints
from Eq. (\ref{eq:Binned mixture}) to arrive at more sensitive criteria.

\begin{figure}
\includegraphics[scale=0.2]{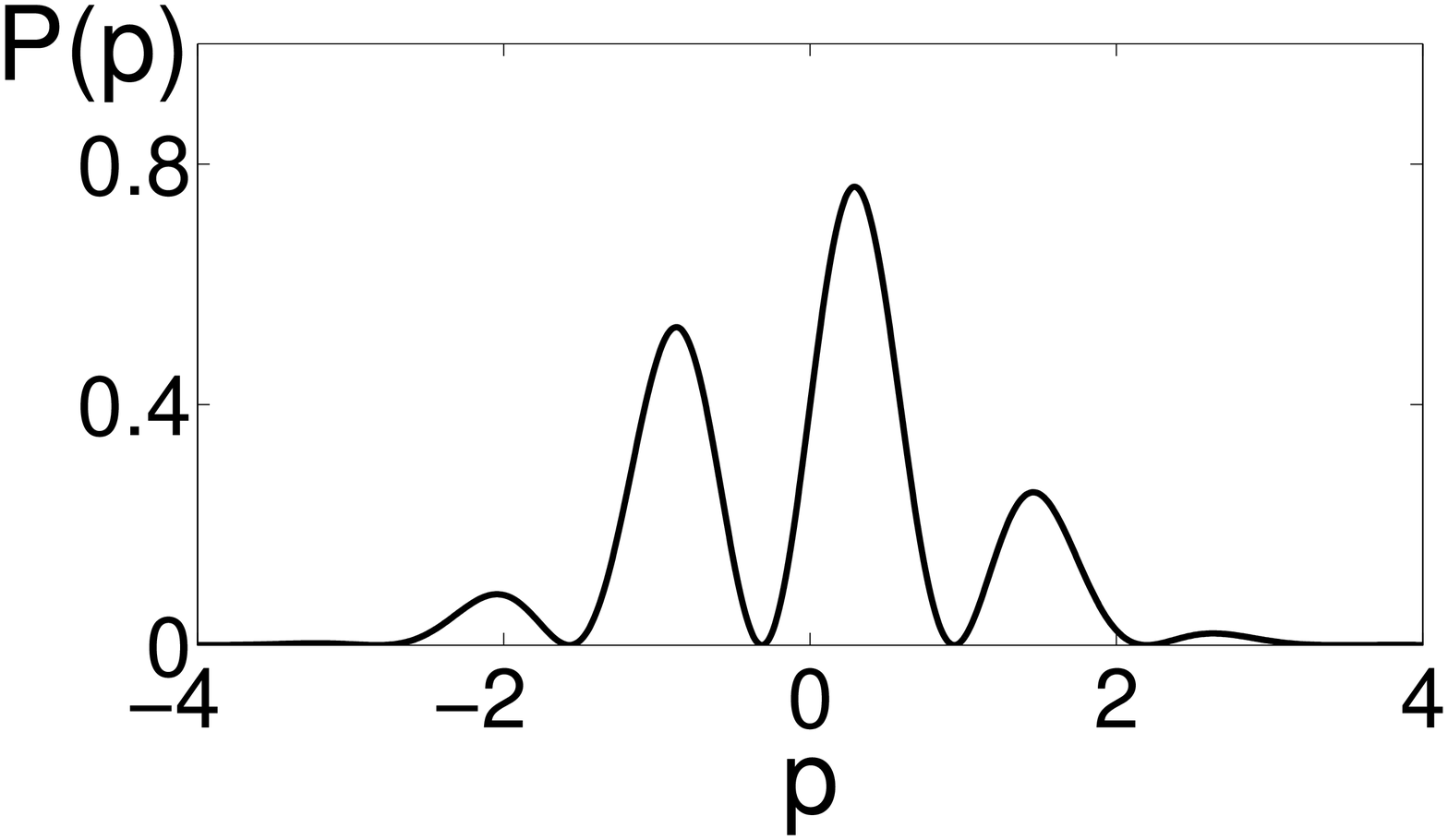}~~\\
 \includegraphics[scale=0.2]{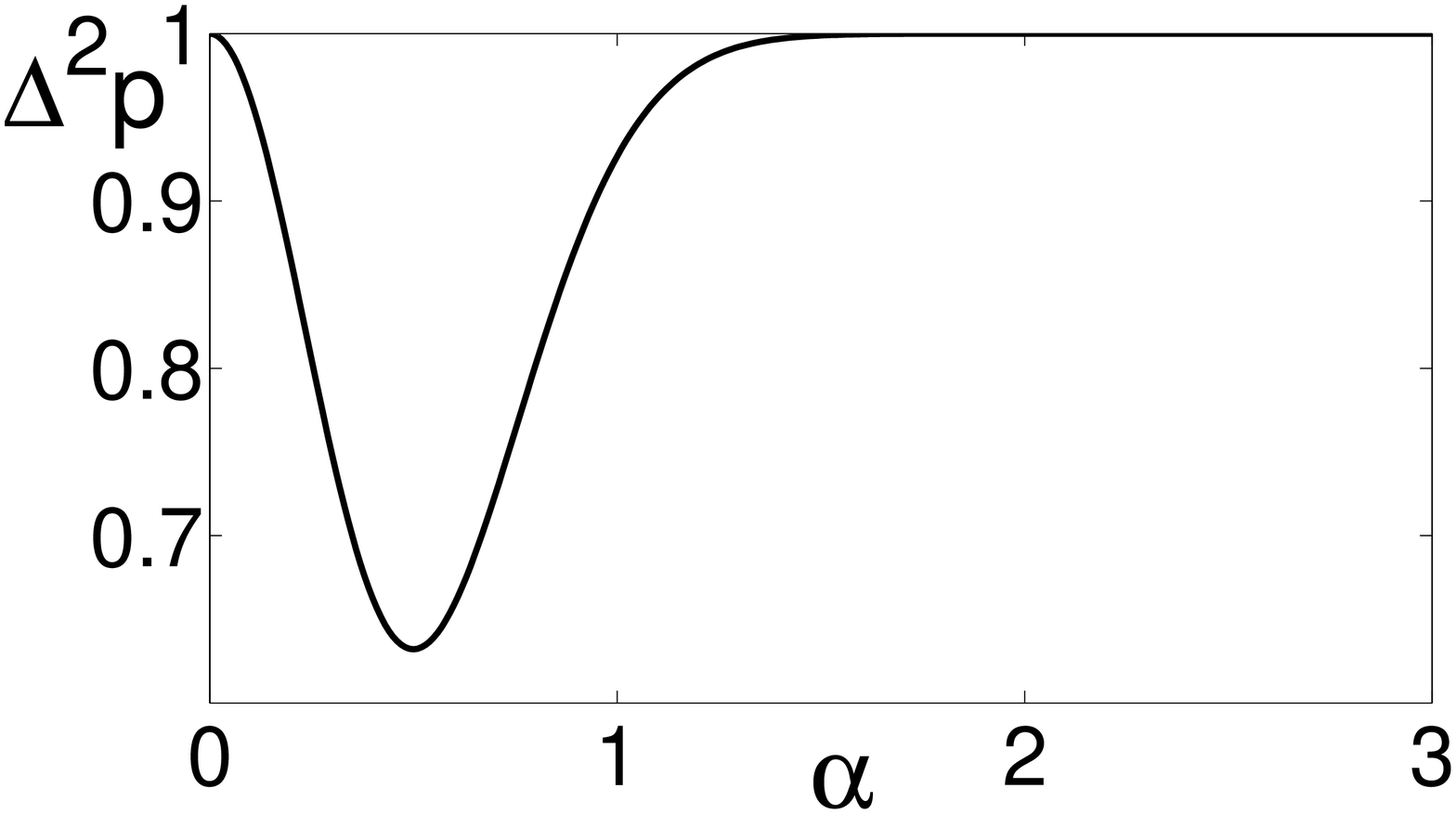}

\caption{(a) $P(p)$ for a superposition (\ref{eq:scatys}) of two coherent
states where $\alpha=2.5$ and (b) the reduced variance $\Delta^{2}p<1$,
versus $\alpha$.}

\end{figure}

To investigate what can be inferred from criteria (\ref{eq:nobinning ineq}),
we note that $\hat{x}$ is the macroscopic observable. The complementary
observable $\hat{p}$ has distribution $P(p)=\exp{[-p^{2}/2]}(1+\sin{2\alpha p)}/\sqrt{2\pi}$
which exhibits fringes and has variance $\Delta^{2}p=1-4\alpha^{2}\exp{[-4\alpha^{2}]}$
(Fig. 8). There is a maximum squeezing of $\Delta^{2}p\approx0.63$
at $\alpha=0.5$. However, the squeezing diminishes as $\alpha$ increases,
so the criterion becomes less effective as the separation of states
of the macroscopic superposition increases. The maximum separation
$S$ that could be conclusively inferred from this criterion is $S\approx2.5$
at $\alpha=0.5$.

As discussed in Sec. VII, the detection of squeezing in $p$ is enough
to confirm the system is \emph{not} that of the mixture \begin{equation}
\rho=1/2(|\alpha\rangle\langle\alpha|+|-\alpha\rangle\langle-\alpha|)\label{eq:mixcoh}\end{equation}
of the two coherent states. In fact, the squeezing rules out that
the system is any mixture of coherent states. We note though that
since the degree of squeezing $\Delta p$ is small, our criteria is
not sensitive enough to rule out superpositions of macroscopically
separated coherent states.

\subsection{Squeezed states}

Consider the single-mode momentum squeezed state \citep{cavesyuen}
\begin{equation}
|\psi\rangle=e^{r(a²-a^{\dagger2})}\left|0\right\rangle \label{eq:sqestate}\end{equation}
 Here $\left|0\right\rangle $ is the vacuum state. For large values
of $r$ these states are generalized macroscopic superpositions of
the continuous set of eigenstates $|x\rangle$ of $\hat{x}=a+a^{\dagger}$,
with wave function\begin{equation}
\left\langle x|\psi\right\rangle =\frac{1}{(2\pi\sigma)^{\frac{1}{4}}}\exp\{\frac{-x^{2}}{4\sigma}\},\label{eq:sqstatewf}\end{equation}
 and associated Gaussian probability distribution \begin{equation}
P(x)=\frac{1}{(2\pi\sigma)^{\frac{1}{2}}}\exp\{\frac{-x^{2}}{2\sigma}\}\label{eq:gausp}\end{equation}
The variance is $\sigma=e^{2r}$. As the squeeze parameter $r$ increases,
the probability distribution expands, so that eventually with large
enough $r$, $x$ can be regarded as a macroscopic observable. This
behavior is shown in Fig. 9. The distribution for $p$ is also Gaussian
but is squeezed, meaning that it has reduced variance: $\Delta^{2}p<1$.
In fact, Eq. (\ref{eq:sqestate}) is a minimum uncertainty state,
with $\Delta^{2}p=1/\sigma=e^{-2r}$. Where squeezing is significant,
the off-diagonal elements $\langle x|\rho|x'\rangle=\langle x|\psi\rangle\langle\psi|x'\rangle$
(where $|x-x'|$ is large) are significant over a large range of $x$
values (Fig. 9). 

\begin{figure}
\includegraphics[scale=0.25]{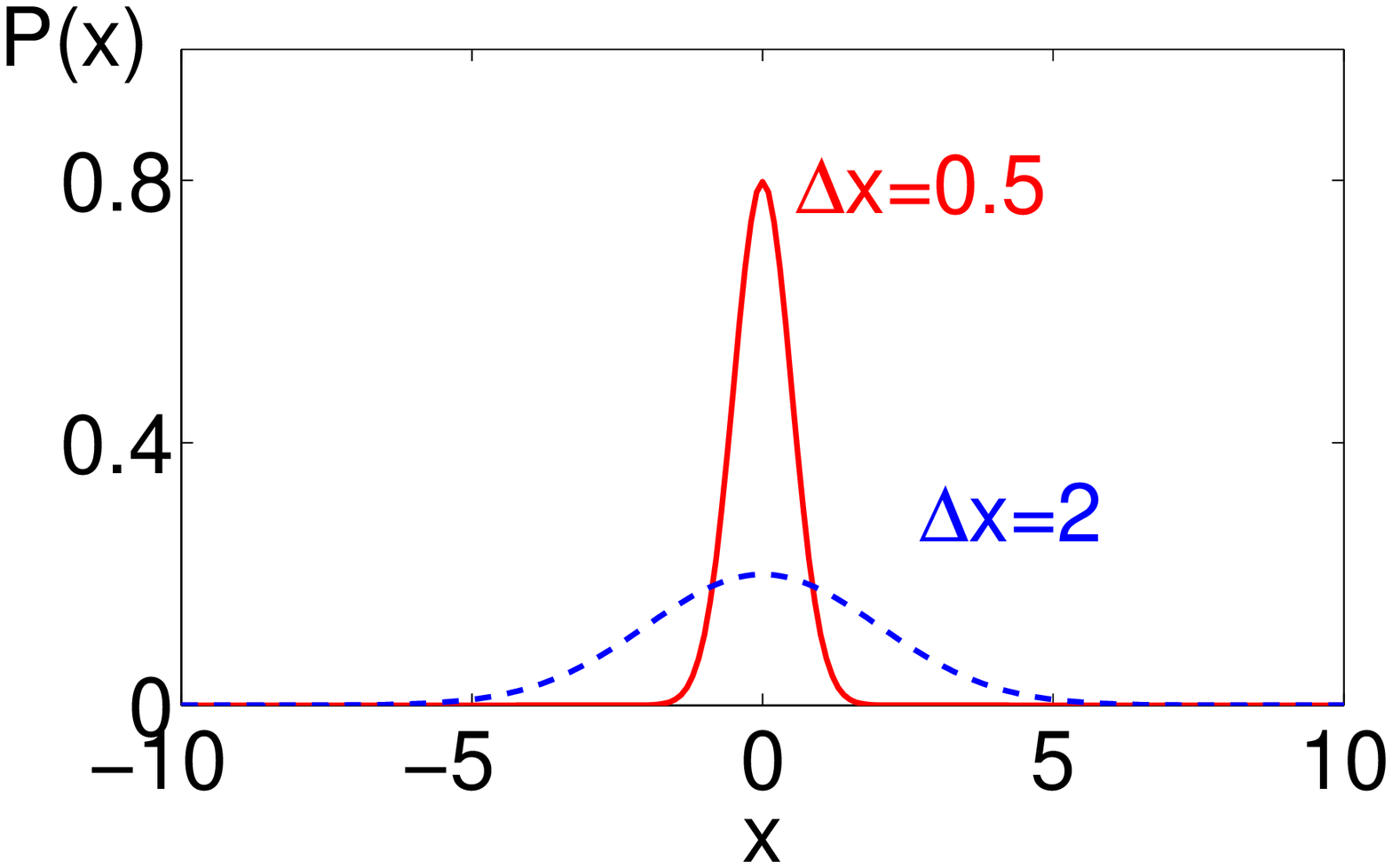}~\\
\includegraphics[scale=0.25]{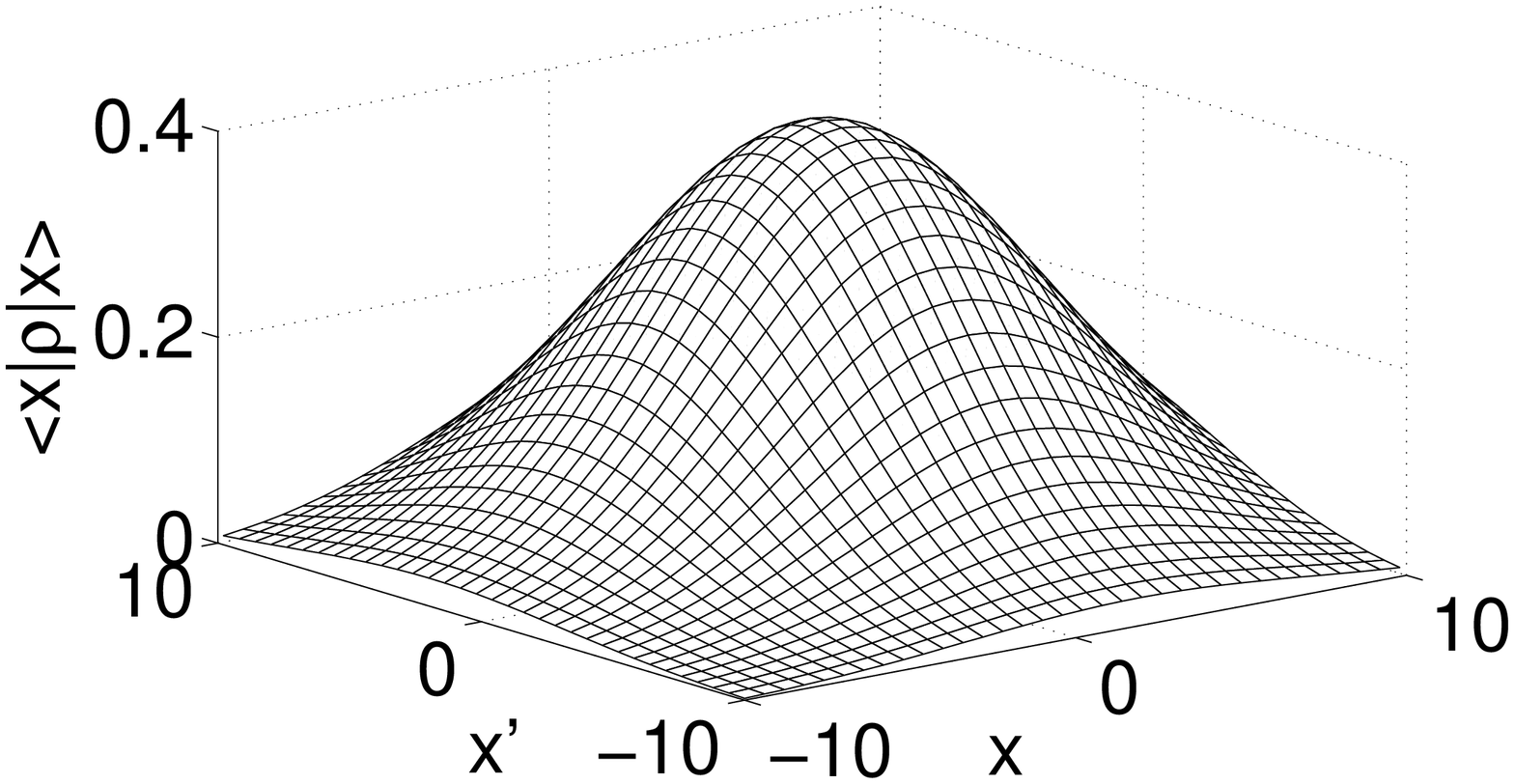}

\caption{(a) Probability distribution for a measurement $X$ for a momentum-squeezed
state. The variance $\Delta^{2}x$ increases with squeezing in $p$,
to give a macroscopic range of outcomes, and for the minimum uncertainty
state (\ref{eq:sqestate}) satisfies $\Delta x\Delta p=1$. (b) The
$\langle x|\rho|x'\rangle$ for a squeezed state (\ref{eq:sqestate})
with $r=13.4$ ($\Delta x=3.67)$ which predicts $\langle a^{\dagger}a\rangle=2.5^{2}$. }

\end{figure}

The criterion (\ref{eq:critsuper}) for the binned outcomes is violated
for the ideal squeezed state (\ref{eq:sqestate}) for values of $S$
up to $0.5\sqrt{\sigma}$. The criterion can thus confirm macroscopic
superpositions of states with separation of up to half the standard
deviation of the probability distribution of $x$, even as $\Delta x\rightarrow\infty$.
This behavior has been reported in \citep{macsupPRL} and is shown
in Fig. 10.

\begin{figure}
\includegraphics[scale=0.28]{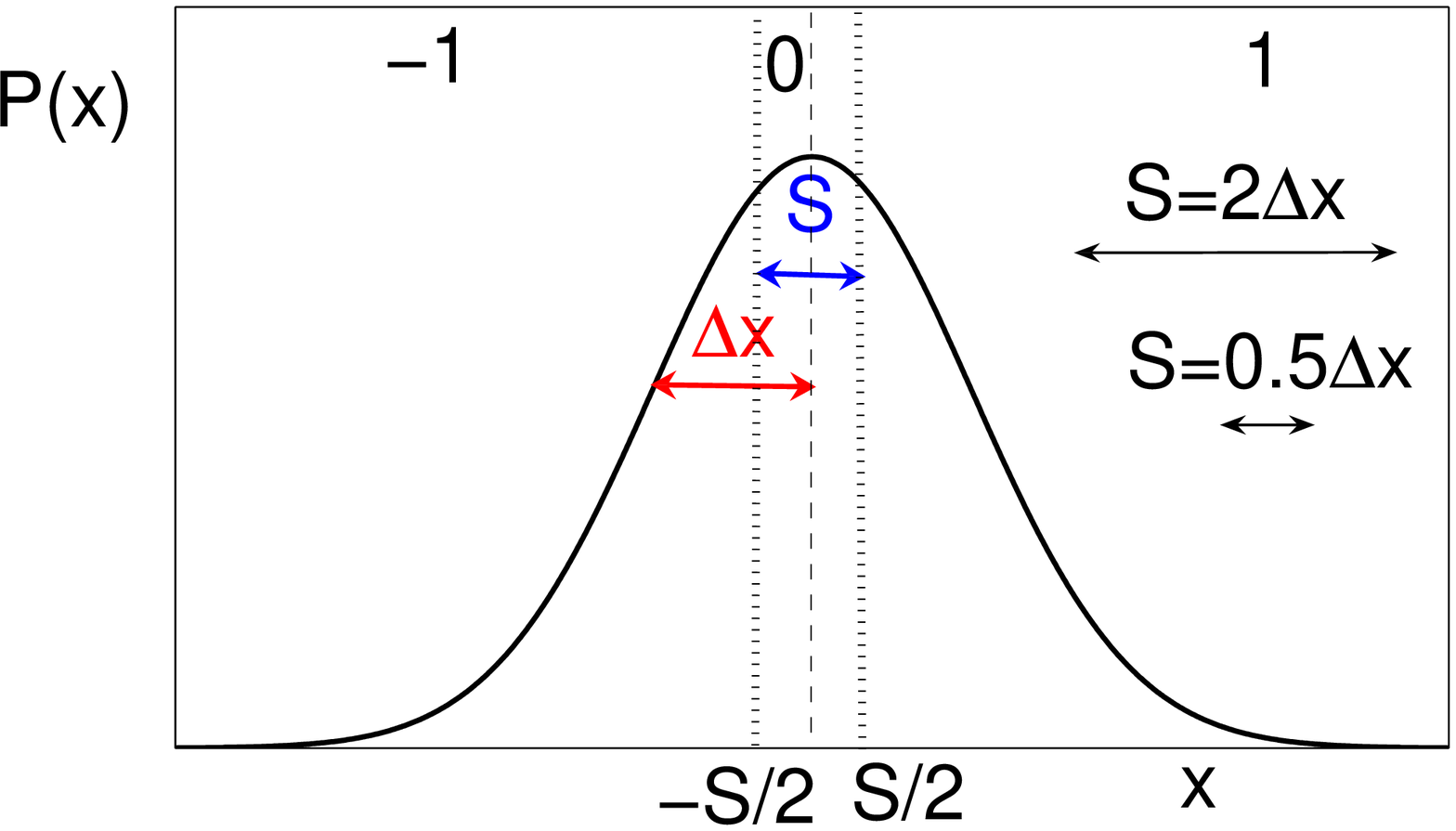}~~\\
 \includegraphics[scale=0.28]{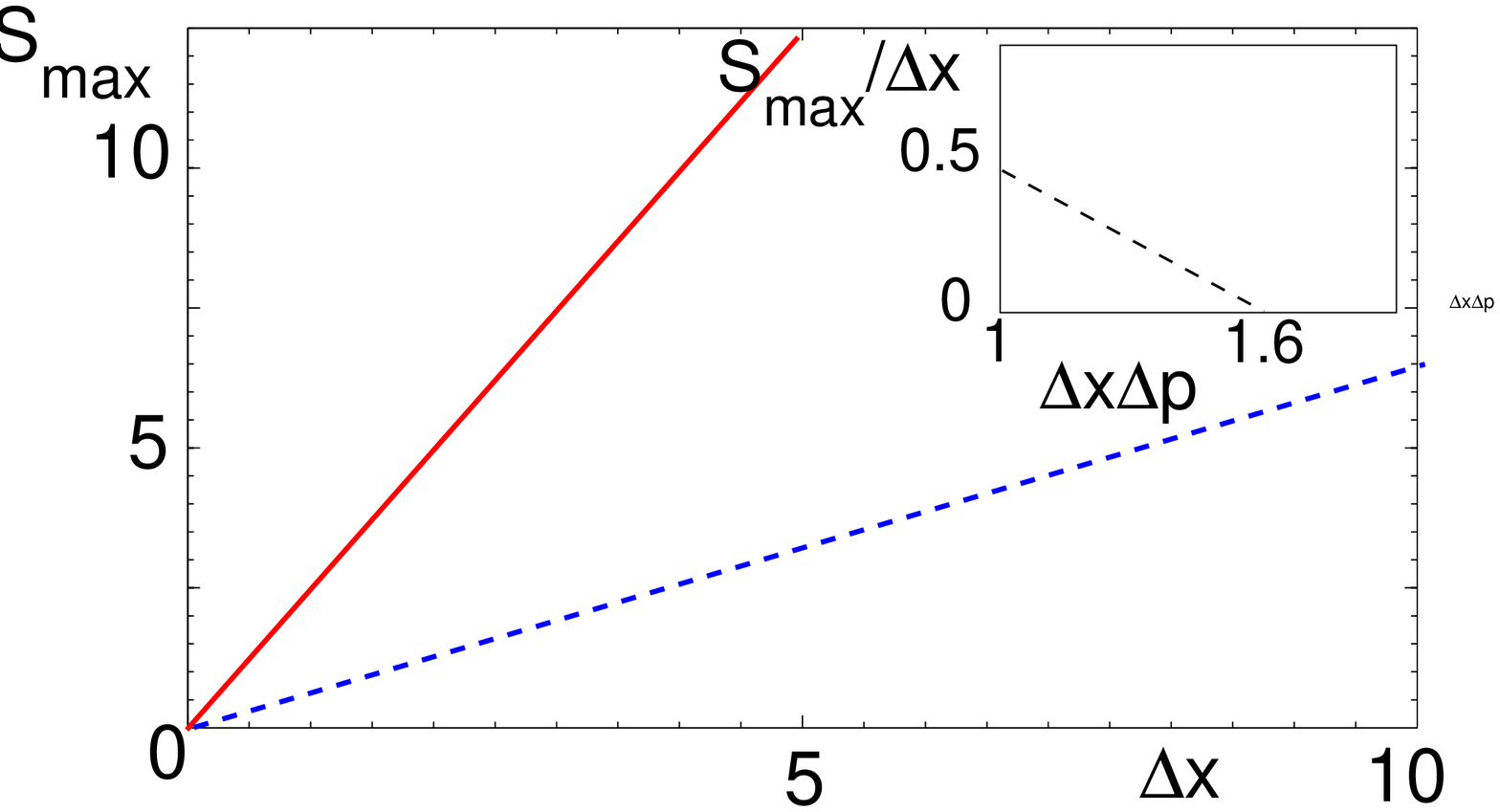}

\caption{Detection of underlying superpositions of size $S$ for the squeezed
minimum uncertainty state (\ref{eq:sqestate}) by violation of (\ref{eq:critsuper})
(dashed line of (b)) and (\ref{eq:nobinning ineq}) (full line of
(b)). $S_{max}$ is the maximum $S$ for which the inequalities are
violated. Inset of (b) shows behavior of violation of (\ref{eq:critsuper})
for general Gaussian-squeezed states. Inequality (\ref{eq:nobinning ineq})
depends only on $\Delta p$. The size of $S_{max}$ relative to $P(x)$
is illustrated in (a). }

\end{figure}

Squeezed systems that are generated experimentally will not be describable
as the pure squeezed state (\ref{eq:sqestate}). This pure state is
a minimum uncertainty state with $\Delta x\Delta p=1$. Typically
experimental data will generate Gaussian probability distributions
for both $x$ and $p$ and with squeezing $\Delta p<1$ in $p$, but
typically $\Delta x\Delta p>1$. The maximum value of $S$ that can
be proved in this case of the Gaussian states reduces to $0$ as $\Delta x\Delta p$
(or $\Delta x\Delta_{inf}p$) increases to $\sim1.6$. This is shown
in Fig. 10. Analysis of recent experimental data for impure states
that allows a violation of Eq. (\ref{eq:critsuper}) has been reported
by Marquardt \emph{et al.} \citep{marqu}.

The criterion (\ref{eq:nobinning ineq}), as given by theorem 4, is
better able to detect the superpositions (Fig. 10), particularly where
the uncertainty product gives $\Delta x\Delta p>1$, though in this
case the superpositions are non-locatable in phase space, so that
we cannot conclude an outcome domain for the states involved in the
superposition. This criterion depends only on the squeezing $\Delta p$
in one quadrature and is not sensitive to the product $\Delta x\Delta p$.
For ideal squeezed states with variance $\Delta^{2}x=\sigma$, one
can prove a superposition of size $S=2\sqrt{\sigma}$, four times
that obtained from Eq. (\ref{eq:critsuper}) (Fig. 10). 

Experimental reports \citep{squresultexp} of squeezing of orders
$\Delta p\approx0.4$ confirms superpositions of size at least $S=5$,
which is 2.5 times that defined by $S=2$, which corresponds to two
standard deviations of the coherent state, for which $\Delta x=1$
(Fig. 4).

\subsection{Two-mode squeezed states}

Next we consider the two-mode squeezed state \citep{cavessch}

\begin{equation}
e^{r(ab-a^{\dagger}b^{\dagger})}|0\rangle|0\rangle\label{eq:twomodesqstate}\end{equation}
Here $a,b$ are boson annihilation operators for modes $A$ and $B$
respectively. The wave function $\langle x|\psi\rangle$ and distribution
$P(x)$ are as in Eqs. (\ref{eq:sqstatewf}) and (\ref{eq:gausp}),
but the variance in $\hat{x}=X^{A}$ is now given by $\sigma=\cosh2r$.
The $\hat{x}=X^{A}$ is thus a macroscopic observable. 

In the two-mode case, the squeezing is in a linear combination $P^{A}+P^{B}$
of the momenta $P^{A}$ and $P^{B}$ at $A$ and $B$, rather than
in the momentum $\hat{p}=P^{A}$ for $A$ itself. The observable that
is complementary to $X^{A}$ is of form $\tilde{P}=P^{A}-gP^{B}$
where $g$ is a constant, which is Eq. (\ref{eq:ptilde}) of Sec.
V. We can select to evaluate one of the criteria (\ref{eqn:eprcat}),
(\ref{eq:nobinning ineq2}) or (\ref{eq:resultcompsum}).

Choosing as our macroscopic observable $x$ and our complementary
one $P^{A}-gP^{B}$, we calculate\begin{equation}
\Delta_{inf}^{2}p^{A}=1/\sigma=1/cosh2r\label{eq:infresult}\end{equation}
for the choice $g=\left\langle P^{A}P^{B}\right\rangle /\left\langle (P^{B})^{2}\right\rangle =-tanhr$
which minimizes $\Delta_{inf}^{2}p^{A}$ \citep{eprr}. The application
of results to criterion (\ref{eqn:eprcat}) gives the result as in
Fig.10, to indicate detection of superpositions of size $S$ where
$S=0.5\sqrt{\sigma}$ for the ideal squeezed state (\ref{eq:twomodesqstate}),
and the result shown in the inset of Fig. 10 if $\Delta x^{A}\Delta_{inf}p^{A}>1$.

The prediction for the criterion of theorem 3, to detect superpositions
in the position sum $X^{A}+X^{B}$ by measurement of a narrowed variance
in the momenta sum $P^{A}+P^{B}$, is also given by the results of
Fig. 10. Calculation for the ideal state (\ref{eq:twomodesqstate})
predicts $\Delta^{2}(\frac{p^{A}+p^{B}}{\sqrt{2}})=e^{-2r}$ and $\Delta^{2}(\frac{x^{A}+x^{B}}{\sqrt{2}})=e^{+2r}$
which corresponds to that of the one-mode squeezed state. The prediction
for the maximum value of $S$ of Theorem 3 is therefore given by the
dashed curves of Fig. 10, and the inset.

A better result is given by Eq. (\ref{eq:nobinning ineq2}), if we
are not concerned with the location of the superposition. Where we
use Eq. (\ref{eq:nobinning ineq2}), the degree of reduction in $\Delta_{inf}^{2}p^{A}$
determines the size of superposition $S$ that may be inferred. By
theorem 5, measurement of $\Delta_{inf}p^{A}$ allows inference of
superpositions of eigenstates of $\hat{x}$ separated by at least
\begin{equation}
S=2/\Delta_{inf}p^{A}\label{eq:sdpinf}\end{equation}
Realistic states are not likely to be pure squeezed states as given
by Eq. (\ref{eq:twomodesqstate}). Nonetheless the degree of squeezing
indicates a size of superposition in $X^{A}$, as given by theorem
5. Experimental values of $\Delta_{inf}^{2}p^{A}\approx0.76$ have
been reported \citep{bowen}, to give confirmation of superpositions
of size $S\approx2.3,$ which is $1.1$ times the level of $S=2$
that corresponds to two standard deviations $\Delta x^{A}=1$ of the
vacuum state (Fig. 4). 

More frequently, it is the practice to measure squeezing in the direct
sum $P^{A}+P^{B}$ of momenta. The macroscopic observable is then
the position sum $X^{A}+X^{B}$. The reports of measured experimental
values indicate \citep{elizclau} $\Delta^{2}(\frac{p^{A}+p^{B}}{\sqrt{2}})\approx0.4$,
which according to theorem 5 implies superpositions in $(X^{A}+X^{B})/\sqrt{2}$
of size $S\approx3.2,$ of order $1.6$ times the standard vacuum
state level. The slightly better experimental result for the superpositions
in the position sum may be understood since it has been shown by Bowen
\emph{et al.} \citep{bowen} that, for the Gaussian squeezed states,
the measurement of $\Delta_{inf}^{2}p^{A}$ is more sensitive to loss
than that of $\Delta^{2}(p^{A}+p^{B})$. The $\Delta_{inf}p^{A}$
is an asymmetric measure that enables demonstration of the EPR paradox
\citep{eprr,mdrepr}, a strong form of quantum nonlocality \citep{eprreview,wisesteer}.

\section{Conclusion}

We have extended our previous work \citep{macsupPRL} and derived
criteria sufficient to detect generalized macroscopic (or $S$-scopic)
superpositions ($\sum_{k_{1}}^{k_{2}}c_{k}|x_{k}\rangle$) of eigenstates
of an observable $\hat{x}$. For these superpositions, the important
quantity is the value $S$ of the \emph{extent} of the superposition,
which is the range in prediction of the observable ($S$ is the maximum
of $|x_{j}-x_{i}|$ where $c_{j},c_{i}\neq0$). This quantity gives
the extent of indeterminacy in the quantum prediction for $\hat{x}$.
In this sense, there is a contrast with the prototype macroscopic
superposition (of type $c_{2}|x_{2}\rangle+c_{1}|x_{1}\rangle$) that
relates directly to the essay of Schrödinger \citep{schroedinger 1935}.
Such a prototype superposition contains only the two states that have
separation $S$ in their outcomes for $x$. Nonetheless, we have discussed
how the generalized superposition is relevant to testing the ideas
of Schrödinger, in that such macroscopic superpositions are shown
to be inconsistent with the hypothesis of a quantum system being in
at most one of two macroscopically separated states.

We have also defined the concept of a generalized $S$-scopic coherence
and the class of minimum uncertainty theories (MUTs) without direct
reference to quantum mechanics. The former is introduced in Sec. IV
as the assumption (\ref{eq:probmacro}) and is associated to the failure
of a generalized assumption of macroscopic reality. This assumption
is that the system is in at most one of two macroscopically distinguishable
states, but that these underlying states are not specified to be quantum
states. The assumption of MUT  is that these component states do at
least satisfy the quantum uncertainty relations. In the derivation
of the criteria of this paper, only two assumptions are made: that
the system does satisfy this generalized macroscopic (\emph{S}-scopic)
reality and that the theory is a MUT . These assumptions lead to inequalities,
which, when violated, generate evidence that at least one of the assumptions
must be incorrect.

We point out that if, in the event of violation of the inequalities,
we opt to conclude the failure of the MUT  assumption, then this does
not imply quantum mechanics to be incorrect, but rather that it is
incomplete, in the sense that the component states can themselves
not be quantum states. It can be said then that violation of the inequalities
of this paper implies at least one of the assumptions of \emph{generalized
macroscopic  (S-scopic) reality} and the\emph{ completeness }of quantum
mechanics\emph{ }is incorrect.

There is a similarity with the Einstein-Podolsky-Rosen argument \citep{epr}.
In the EPR argument, the assumption of a form of realism (local realism)
is shown to be inconsistent with the completeness of quantum mechanics.
Therefore, as a conclusion of that argument, one is left to conclude
that at least one of \emph{local realism} and the \emph{completeness
of QM} is incorrect \citep{mdrepr,wiseepr,eprreview}. EPR opted for
the first and took their argument as a demonstration that quantum
mechanics was incomplete. Only after Bell \citep{bell} was it shown
that this was an incorrect choice. Here, as in the EPR argument, the
assumption of a form of realism {[}macroscopic ( \emph{S-}scopic realism]
can only be made consistent with the predictions of quantum mechanics
if one allows a kind of theory in which the underlying states are
not restricted by the uncertainty relations \citep{macsupPRL}.

\begin{acknowledgments}
We thank C. Marquardt, P. Drummond, H. Bachor, Y-C. Liang, N. Menicucci,
N. Korolkova, A. Caldeira, P. K. Lam, H. Wiseman, A. Bradley, M. Olsen,
F. De Martini, E. Giacabino, B. Whaley, G. Leuchs, C. Fabre, A. Leggett,
L. Plimak and others for interesting discussions. We are grateful
for support from the ARC Centre of Excellence Program and the Queensland
State Government.
\end{acknowledgments}
\textbf{APPENDIX A: PROOF OF THEOREM A}

We will now prove the statement that coherence between $x_{1}$ and
$x_{2}$ is equivalent to a nonzero off-diagonal element $\langle x_{1}|\rho|x_{2}\rangle$
in the density matrix. As discussed in Sec. \ref{sec:Generalized-S-scopic-Coherence},
within quantum mechanics the statement that there exists coherence
between $x_{1}$ and $x_{2}$ is equivalent to the statement that
there is no decomposition of the density matrix of form (\ref{eq:QM mixture})
where $\rho_{1}$ and $\rho_{2}$ are density matrices such that $\langle x_{1}|\rho_{2}|x_{1}\rangle=\langle x_{2}|\rho_{1}|x_{2}\rangle=0$.
Therefore theorem A can be reformulated as saying that $\langle x_{1}|\rho|x_{2}\rangle=0$
if and only if such a decomposition \emph{does} exist.

It is easy to prove the first direction of the equivalence: if $\exists$$\{\wp_{1},\,\wp_{2},\,\rho_{1},\,\rho_{2}\}$
such that $\rho=\wp_{1}\rho_{1}+\wp_{2}\rho_{2}$ and $\langle x_{1}|\rho_{2}|x_{1}\rangle=\langle x_{2}|\rho_{1}|x_{2}\rangle=0$,
then $\langle x_{1}|\rho|x_{2}\rangle=0$. To show this, first note
that for any density matrix $\bar{\rho}$ and $\forall\,\{x,x'\}$,
if $\langle x|\bar{\rho}|x\rangle=0$ then $\langle x|\bar{\rho}|x'\rangle=0$
, where $\langle x|x'\rangle=\delta_{x,x'}$. Since by assumption
$\langle x_{1}|\rho_{2}|x_{1}\rangle=\langle x_{2}|\rho_{1}|x_{2}\rangle=0$,
then $\langle x_{1}|\rho|x_{2}\rangle=\sum_{i}\wp_{i}\langle x_{1}|\rho_{i}|x_{2}\rangle=0$.

The converse can also be proved. We use the facts that any $\rho$
can always be written as the reduced density matrix of an enlarged
pure state, where the system of interest (call it $A$) is entangled
with an ancilla $B$, i.e $\rho=Tr_{B}\{|\Psi_{AB}\rangle\langle\Psi_{AB}|\};$
and that any bipartite pure state can always be written in the Schmidt
decomposition \citet{schmidtdec}\begin{equation}
|\Psi_{AB}\rangle=\sum_{i}\sqrt{\eta_{i}}|\psi_{i}\rangle|\phi_{i}^{B}\rangle.\label{eq:purification}\end{equation}
where $\{|\psi_{i}\rangle\}$ and $\{|\phi_{i}^{B}\rangle\}$ are
orthonormal and $\eta_{i}\in[0,1].$ The superscript $B$ denotes
the states of the ancilla and the absence of a superscript denotes
the states of the system of interest, $A$. We decompose each pure
state $|\psi_{i}\rangle$ that appears in the Schmidt decomposition
in the basis of eigenstates of $\hat{x}$ as $|\psi_{i}\rangle=\sum_{k}c_{i,k}|x_{k}\rangle$.
By assumption $\langle x_{1}|\rho|x_{2}\rangle=0$ and therefore $\sum_{i}\eta_{i}\langle x_{1}|\psi_{i}\rangle\langle\psi_{i}|x_{2}\rangle=\sum_{i}\eta_{i}c_{i,1}c_{i,2}^{*}=0$.
We can expand $|\Psi_{AB}\rangle$ as\begin{equation}
|\Psi_{AB}\rangle=|x_{1}\rangle|\tilde{1_{B}}\rangle+|x_{2}\rangle|\tilde{2_{B}}\rangle+\sum_{k>2,\, i}\sqrt{\eta_{i}}c_{i,k}|x_{k}\rangle|\phi_{i}^{B}\rangle,\label{eq:purification expansion}\end{equation}
where we define the (unnormalized) $|\tilde{1_{B}}\rangle\equiv\sum_{i}\sqrt{\eta_{i}}c_{i,1}|\phi_{i}^{B}\rangle$
and $|\tilde{2_{B}}\rangle\equiv\sum_{i}\sqrt{\eta_{i}}c_{i,2}|\phi_{i}^{B}\rangle$.
The inner product of these two vectors is $\langle\tilde{1_{B}}|\tilde{2_{B}}\rangle=\sum_{i}\eta_{i}c_{i,1}c_{i,2}^{*}.$
But as shown above $\sum_{i}\eta_{i}c_{i,1}c_{i,2}^{*}=0$, so $|\tilde{1_{B}}\rangle$
and $|\tilde{2_{B}}\rangle$ are orthogonal. We can therefore define
an orthonormal basis with the (normalized) $|1_{B}\rangle=|\tilde{1_{B}}\rangle/\sqrt{\sum_{i}\eta_{i}|c_{i,1}|^{2}}$
and $|2_{B}\rangle=|\tilde{2_{B}}\rangle/\sqrt{\sum_{i}\eta_{i}|c_{i,2}|^{2}},$
plus additional $|j_{B}\rangle$with $3\leq j\leq D$, where $D$
is the dimension of subsystem $B$'s Hilbert space. Taking the trace
of $\rho_{AB}=|\Psi_{AB}\rangle\langle\Psi_{AB}|$ therefore yields\begin{eqnarray}
\rho & = & Tr_{B}\{\rho_{AB}\}\nonumber \\
 & = & \left\langle 1_{B}\right|\rho_{AB}\left|1_{B}\right\rangle +\left\langle 2_{B}\right|\rho_{AB}\left|2_{B}\right\rangle \nonumber \\
 &  & +\sum_{j>2}\left\langle j_{B}\right|\rho_{AB}\left|j_{B}\right\rangle .\label{eq:traceB}\end{eqnarray}
Now referring to expansion \eqref{eq:purification expansion}, we
see that $\langle1_{B}|\rho_{AB}|1_{B}\rangle=\sum_{i}\eta_{i}|c_{i,1}|^{2}|x_{1}\rangle\langle x_{1}|$
and $\langle2_{B}|\rho_{AB}|2_{B}\rangle=\sum_{i}\eta_{i}|c_{i,2}|^{2}|x_{2}\rangle\langle x_{2}|.$
We then define $\rho_{1}\equiv|x_{1}\rangle\langle x_{1}|,$ $\wp_{1}\equiv\sum_{i}\eta_{i}|c_{i,1}|^{2},$
$\wp_{2}=1-\wp_{1}$ and $\rho_{2}\equiv\frac{1}{\wp_{2}}\{\sum_{i}\eta_{i}|c_{i,2}|^{2}|x_{2}\rangle\langle x_{2}|+\sum_{j>2}\langle j_{B}|\rho_{AB}|j_{B}\rangle\}.$
Obviously $\langle x_{2}|\rho_{1}|x_{2}\rangle=0,$ and by substituting
Eq. \eqref{eq:purification expansion} into $\rho_{2}$ we see that\textbf{
}$\langle x_{1}|\rho_{2}|x_{1}\rangle=0.$ Therefore $\rho$ can be
decomposed as $\rho=\wp_{1}\rho_{1}+\wp_{2}\rho_{2}$ with $\langle x_{1}|\rho_{2}|x_{1}\rangle=\langle x_{2}|\rho_{1}|x_{2}\rangle=0$
as desired.

\textbf{APPENDIX B}

We wish to prove that if $\rho$ can be written as $\rho_{mix}=\wp_{L}\rho_{L}+\wp_{R}\rho_{R}$,
then $\Delta_{inf,mix}^{2}p^{A}\geq\wp_{L}\Delta_{inf,L}^{2}p^{A}+\wp_{R}\Delta_{inf,R}^{2}p^{A}$,
where \[
\Delta_{inf,J}^{2}p^{A}=\sum_{p^{B}}\wp_{J}(p^{B})\Delta_{J}^{2}(p^{A}|p^{B}).\]
The subscript $J$ refers to the $\rho_{J}$ from which the probabilities
are calculated.

We have\begin{eqnarray*}
\Delta_{inf,mix}^{2}p^{A} & = & \sum_{p^{B}}P_{mix}(p^{B})\Delta_{mix}^{2}(p^{A}|p^{B})\\
 & = & \sum_{p^{B}}\sum_{p^{A}}P_{mix}(p^{A},p^{B})(p^{A}-\langle p^{A}|p^{B}\rangle_{mix})^{2}\\
 & \geq & \sum_{p^{B}}\sum_{p^{A}}\sum_{I=R,L}\wp_{I}P_{I}(p^{A},p^{B})(p^{A}-\langle p^{A}|p^{B}\rangle_{I})^{2}\end{eqnarray*}
The inequality follows because $\langle p^{A}|p^{B}\rangle_{mix}$
is the mean of $P(p^{A}|p^{B})$ for $\rho_{mix}$, and the choice
$a=\sum_{p}P(p)p=\langle p\rangle$ will minimize $\sum_{p}P(p)(p-a)^{2}$.
From this the required result follows.

\end{document}